\documentclass[journal]{IEEEtran}

\usepackage{subfigure}
\usepackage{times}
\usepackage{epsfig}
\usepackage{graphicx}
\usepackage{amsmath}
\usepackage{amssymb}
\usepackage{multirow}
\usepackage{color}
\usepackage{cite}
\usepackage{array}
\begin{document}
%
\title{Deep Learning for Hyperspectral Image Classification: An Overview}
\author{Shutao Li,~\IEEEmembership{Fellow,~IEEE,} Weiwei Song,~\IEEEmembership{Student Member,~IEEE,} Leyuan Fang,~\IEEEmembership{Senior Member,~IEEE,}  Yushi Chen,~\IEEEmembership{Member,~IEEE,} Pedram Ghamisi,~\IEEEmembership{Senior Member,~IEEE,} and
J\'on Atli Benediktsson,~\IEEEmembership{Fellow,~IEEE}
\thanks{ This work was supported by the National Natural Science Fund of China under Grant 61890962, 61520106001 and 61771192, the Science and Technology Plan Project Fund of Hunan Province under Grant CX2018B171, 2017RS3024 and 2018TP1013, the Science and Technology Talents Program of Hunan Association for Science and Technology under Grant 2017TJ-Q09, and the National Key RD Program of China under Grant 2018YFB1305200.}

\thanks{S. Li, W. Song, and L. Fang are with the College of Electrical and Information Engineering, Hunan University, Changsha, 410082, China, and also with the Key Laboratory of Visual Perception and Artificial Intelligence of Hunan Province, Changsha, 410082, China (e-mail: shutao\_li@hnu.edu.cn; weiwei\_song@hnu.edu.cn; fangleyuan@gmail.com). \par
Y. Chen is with the Department of Information Engineering, School of Electronics and Information Engineering, Harbin Institute of Technology, Harbin 150001, China (e-mail: chenyushi@hit.edu.cn). \par
P. Ghamisi is with the Helmholtz-Zentrum Dresden-Rossendorf (HZDR), Helmholtz Institute Freiberg for Resource Technology (HIF), Exploration, D-09599 Freiberg, Germany (e-mail: p.ghamisi@gmail.com).  \par
J. A. Benediktsson is with the Faculty of Electrical and Computer Engineering,
University of Iceland, 101 Reykjavk, Iceland (e-mail: benedikt@hi.is).}
}


\maketitle

\begin{abstract}

Hyperspectral image (HSI) classification has become a hot topic in the field of remote sensing. In general, the complex characteristics of hyperspectral data make the accurate classification of such data challenging for traditional machine learning methods. In addition, hyperspectral imaging often deals with an inherently nonlinear relation between the captured spectral information and the corresponding materials. In recent years, deep learning has been recognized as a powerful feature-extraction tool to effectively address nonlinear problems and widely used in a number of image processing tasks. Motivated by those successful applications, deep learning has also been introduced to classify HSIs and demonstrated good performance. This survey paper presents a systematic review of deep learning-based HSI classification literatures and compares several strategies for this topic. Specifically, we first summarize the main challenges of HSI classification which cannot be effectively overcome by traditional machine learning methods, and also introduce the advantages of deep learning to handle these problems. Then, we build a framework which divides the corresponding works into spectral-feature networks, spatial-feature networks, and spectral-spatial-feature networks to systematically review the recent achievements in deep learning-based HSI classification. In addition, considering the fact that available training samples in the remote sensing field are usually very limited and training deep networks require a large number of samples, we include some strategies to improve classification performance, which can provide some guidelines for future studies on this topic. Finally, several representative deep learning-based classification methods are conducted on real HSIs in our experiments.

\end{abstract}


%
\IEEEpeerreviewmaketitle

\section{Introduction}

\IEEEPARstart{H}{yperspectral} imaging is an important technique in remote sensing, which collects the electromagnetic spectrum from the visible to the near-infrared wavelength ranges. Hyperspectral imaging sensors often provide hundreds of narrow spectral bands from the same area on the surface of the earth. In hyperspectral images (HSIs), each pixel can be regarded as a high-dimensional vector whose entries correspond to the spectral reflectance in a specific wavelength. With the advantage of distinguishing subtle spectral difference, HSIs have been widely applied in many fields \cite{agriculture,environment,mineral,Dian-HSI-MSI, Sparse_Unmixing}.

Based on the recent studies published in \cite{Pedram-HSI-SP-review}, HSI classification (i.e., assigning each pixel to one certain class based on its spectral characteristics) is the most vibrant field of research in the hyperspectral community and has drawn broad attentions in the remote sensing field. In HSI classification tasks, there exist two main challenges: 1) the large spatial variability of spectral signatures and 2) the limited available training samples versus the high dimensionality of hyperspectral data. The first challenge is often brought by many factors such as changes in illumination, environmental, atmospheric, and temporal conditions. The second challenge will result in ill-posed problems for some methods and reduce the generalization ability of classifiers.   \par

In the early stage of the study on HSI classification, most methods have focused on exploring the role of the spectral signatures of HSIs for the purpose of classification. Thus, numerous pixel-wise classification methods (e.g., neural networks \cite{NN}, support vector machines (SVM) \cite{SVM}, multinomial logistic regression \cite{MLR1}, \cite{MLR2}, and dynamic or random subspace \cite{RF1}, \cite{RF2}) have been proposed to classify HSIs. In addition, some other classification approaches have focused on designing an effective feature-extraction or dimension-reduction technique, such as principle component analysis (PCA) \cite{PCA1}, \cite{PCA2}, independent component analysis (ICA) \cite{ICA}, and linear discriminant analysis (LDA) \cite{LDA}. However, the classification maps obtained by these pixel-wise classifiers are unsatisfactory since the spatial contexts are not considered. Recently, spatial features have been reported to be very useful in improving the representation of hyperspectral data and increasing the classification accuracies \cite{Pedram-HSIC-review, Li-HSIC-review}. More and more spectral-spatial features-based classification frameworks have been developed, which incorporates the spatial contextual information into pixel-wise classifiers. For example, in \cite{EMP1}, \cite{EMP2}, extended morphological profiles (EMPs) were used to exploit the spatial information via multiple morphological operations. Multiple kernel learning (e.g., composite kernel \cite{CK-SVM} and morphological kernel \cite{MK-SVM, SC-MK}) was designed to explore the spectral-spatial information of HSIs. In \cite{EPF}, edge-preserving filtering was considered as a postprocessing technique to optimize the probabilistic results of an SVM. In \cite{JSR, MASR, MF-ASR}, the spatial information within a neighboring region was incorporated into a sparse representation model. These sparse representation methods are based on the observation that hyperspectral pixels can usually be represented by a linear combination of a few common pixels from the same class. Furthermore, spatial consistency was explored by segmenting HSIs into multiple superpixels based on the similarity of either intensity or texture \cite{pixel-superpixel, subpixel-superpixel, EP}. Although most of spectral-spatial-based HSI classification methods have obtained good performance, they heavily depend on hand-crafted or shallow-based descriptors. However, most hand-crafted features are usually designed for a specific task and depend on expert knowledge in the parameter setup phase, which limits the applicability of those approaches in difficult scenarios. The ability of representation of hand-crafted features may not be enough to discriminate subtle variation between the different classes or a large variation between the same classes. Extracting more discriminative features is considered as a crucial procedure for HSI classification.  \par

\begin{figure}
\begin{center}
\includegraphics[width=85mm]{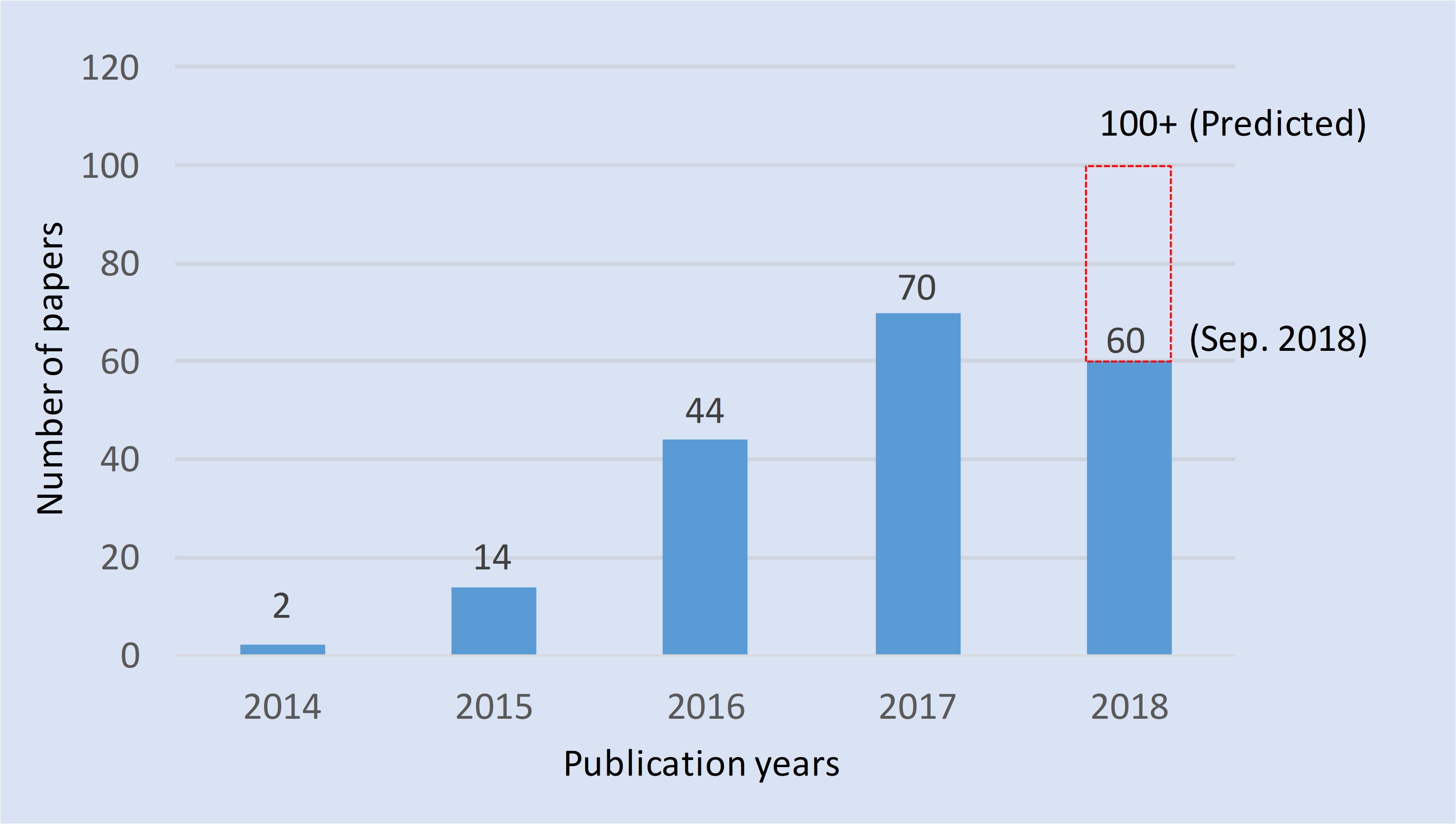}
\end{center}
\caption{The statistics for published papers related to deep learning in HSI classification according to the web of science.}
\label{statistics}
\end{figure}

Recently, deep learning has become a growing trend in big data analysis and great breakthrough has been made with the approach in many computer vision tasks, e.g., image classification \cite{Alexnet, googlenet}, object detection \cite{RCNN}, and natural language processing \cite{NLP}. Motivated by those successful applications, deep learning has been introduced to classify HSIs and achieved good performance. Fig. \ref{statistics} demonstrates the statistics for published papers related to deep learning-based HSI classification in the past five years according to the web of science\footnote{http://http://www.webofknowledge.com/WOS}. From this figure, we can conclude that the topic will be further explored with deep learning and more and more research works will be published in the next several years. Compared with traditional hand-crafted methods, deep learning techniques can extract informative features from the original data via a series of hierarchical layers. Specifically, earlier layers extract some simple features like texture and edge information. Furthermore, the deeper layers are able to represent more complicated features. The learning process is totally automatic, which makes deep learning more suitable for coping with the varieties of situations. In general, deep learning is recognized as an effective feature extraction approach during HSI classification while different networks focus on extracting different feature types.   \par

In this paper, we focus on deep learning-based classification methodologies for HSIs and aim at providing a relatively general and comprehensive overview on the existing methods. The motivation of our work is based on two aspects. On one hand, this work is expected to clarify the mechanism behind existing deep learning-based classification methods. We systematically review a large number of relevant papers in the literature within the framework, where the papers are divided into spectral-feature networks, spatial-feature networks, and spectral-spatial-feature networks with respect to the types of features extracted by the adopted deep networks. On the other hand, we intend to include some strategies to handle the problem of limited available samples (i.e., the second challenge mentioned above), which is very important for designing a deep-learning method for HSI classification.   \par

The remainder of this paper is organized as follows: Section II introduces several typical deep models. In Section III, we categorize the previous works into three aspects with respect to the types of features extracted by deep networks. Some strategies to cope with the problem of limited available samples are introduced in Section IV. In Section V, several representative deep learning-based classification methods are compared on real HSIs. Finally, conclusions and suggestions are provided in Section VI.



\section{Deep Models}
In this section, we briefly introduce several deep network models that have been widely used in the HSI classification field. These deep networks include stacked auto-encoders (SAEs), deep belief networks (DBNs), convolutional neural networks (CNNs), recurrent neural networks (RNNs), and generative adversarial networks (GANs).  \par

\subsection{SAEs}

\begin{figure}
 \centering
  \includegraphics[width=0.8\linewidth]{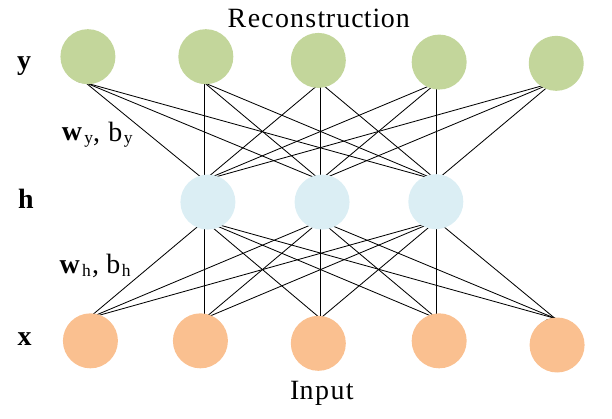}
 \caption{A single hidden-layer auto-encoder. The model learns a hidden feature ``$\mathbf{h}$" by minimizing the error between the input ``$\mathbf{x}$" and the reconstructed output ``$\mathbf{y}$" \cite{Ghamisi-review-2017}.}
 \label{fig:SAE}
\end{figure}

Auto-encoder (AE) is known as the main building block of the stacked AE (SAE) \cite{Chen-SAE}. Fig.~\ref{fig:SAE} shows an AE composing of one visible layer of \textit{d} inputs, one hidden layer of \textit{L} units, and one reconstruction layer of \textit{d} units.
The training procedure has two steps. First, $\mathbf{x} \in {\rm I\!R}^d$ is mapped to $\mathbf{h} \in {\rm I\!R}^L$ in the hidden layer (i.e., ``encoder"). Second, $\mathbf{h}$ is mapped to $\mathbf{y} \in {\rm I\!R}^d$  (i.e., ``decoder"). These two steps can also be defined as follows:
\[
\mathbf{h} = f(\mathbf{w}_h \mathbf{x} +b_h )
\]
\[
\mathbf{y}=f(\mathbf{w}_y \mathbf{x}+b_y )
\]
where $\mathbf{w}_h$ and $\mathbf{w}_y$ denote the input-to-hidden and the hidden-to-output weights, respectively. $b_h$ and $b_y$ represent the \textit{bias} of the hidden and output units, and $f(\cdot)$ represents an activation function. The reconstruction error is estimated using the Euclidean distance between $\mathbf{x}$ and $\mathbf{y}$ to approximate the input data $\mathbf{x}$ by minimizing $\left\| \mathbf{x} - \mathbf{y}\right\|_2^2$.

\begin{figure}
 \centering
  \includegraphics[width=1\linewidth]{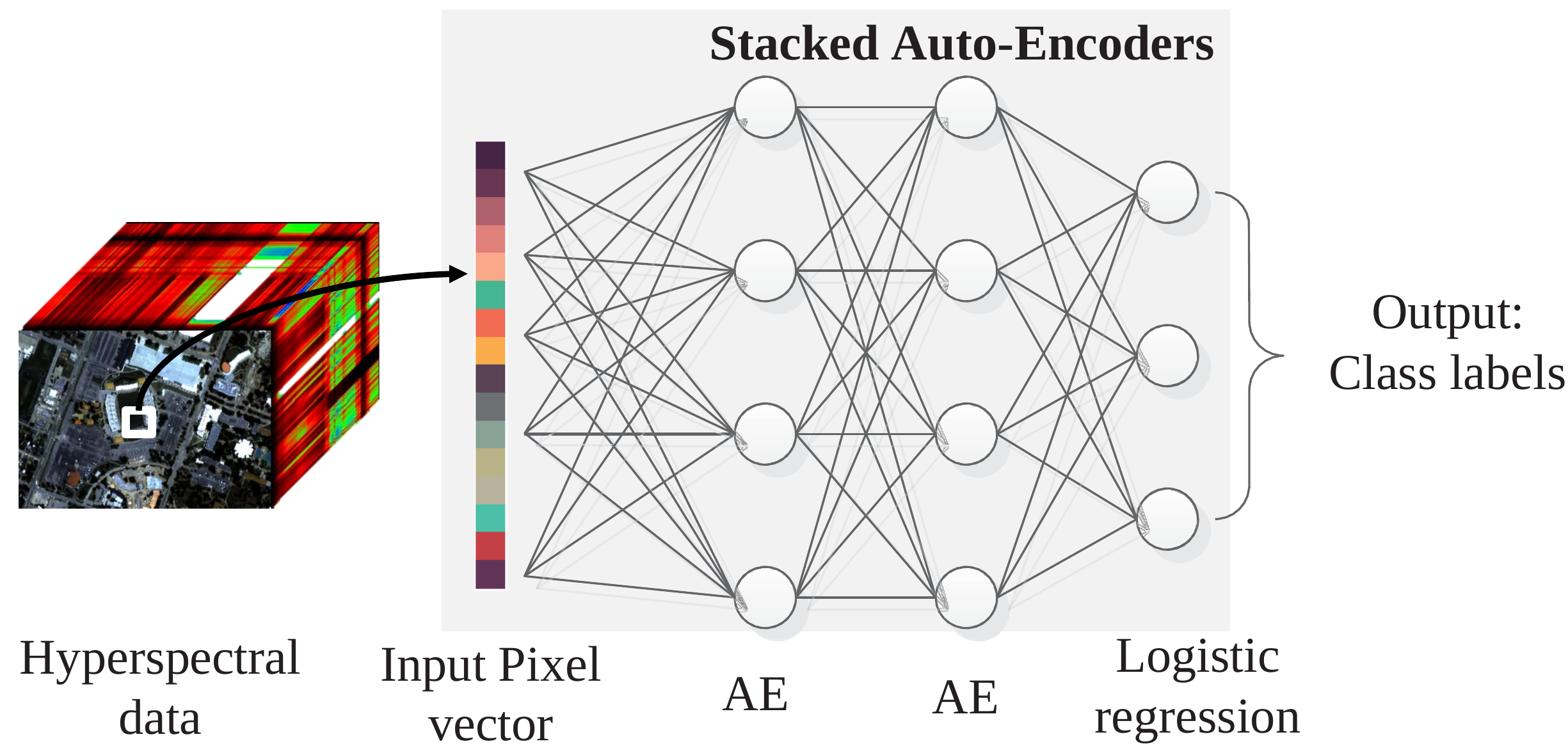}
 \caption{A general illustration of an SAE connected with a subsequent logistic regression classifier \cite{Ghamisi-review-2017}.}
 \label{fig:SAE2}
\end{figure}

SAE can be constructed by stacking multiple layers of AEs that attaches the output of one layer to the input of the following layers. Fig. \ref{fig:SAE2} illustrates a simple representation of an SAE connected with a subsequent logistic regression classifier. The SAE can be used as a spectral classifier where each pixel vector can be considered as input.

\subsection{DBNs}

\begin{figure}
 \centering
  \includegraphics[width=0.8\linewidth]{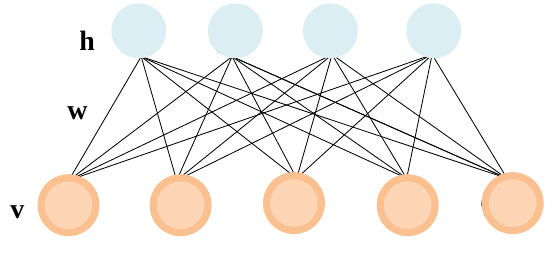}
 \caption{Graphical illustration of a restricted Boltzmann machine. The top layer represents the hidden units and the bottom layer represents the visible units \cite{Ghamisi-review-2017}.}
 \label{fig:DBN1}
\end{figure}

Restricted Boltzmann machine (RBM) is a layer-wise training model and considered as the main building block of a DBN \cite{Chen-DBN}. Fig. \ref{fig:DBN1} illustrates an RBM, consisting of a two-layer network with ``visible" units $\mathbf{v}=\left\{0,1\right\}^d$  and ``hidden" units $\mathbf{h}=\left\{0,1\right\}^L$. Given these units, the energy of a joint configuration of the units can be defined by:
\begin{align}
E(\mathbf{v},\mathbf{h};\theta)=-\sum_{i=1}^d b_i v_i - \sum_{j=1}^L a_j h_j - \sum_{i=1}^d \sum_{j=1}^L w_{ij} v_i h_j
\end{align}
\[
=-\mathbf{b}^\mathbf{T} \mathbf{v}-\mathbf{a}^\mathbf{T} \mathbf{h}-\mathbf{v}^\mathbf{T} \mathbf{wh}
\]
where $\theta=\left\{b_i,a_j,w_{ij} \right\}$. Here, $w_{ij}$ is the weight between the visible unit \textit{i} and hidden unit \textit{j}. $b_i$ and $a_j$ represent the \textit{bias} of the visible and hidden units, respectively. The weight, $w_{ij}$, is learned using the divergence \cite{Chen-DBN}. In RBM, the hidden units are conditionally independent given the visible states, and consequently, can obtain an unbiased sample from the posterior distribution when given a data vector.

\begin{figure}
 \centering
  \includegraphics[width=1\linewidth]{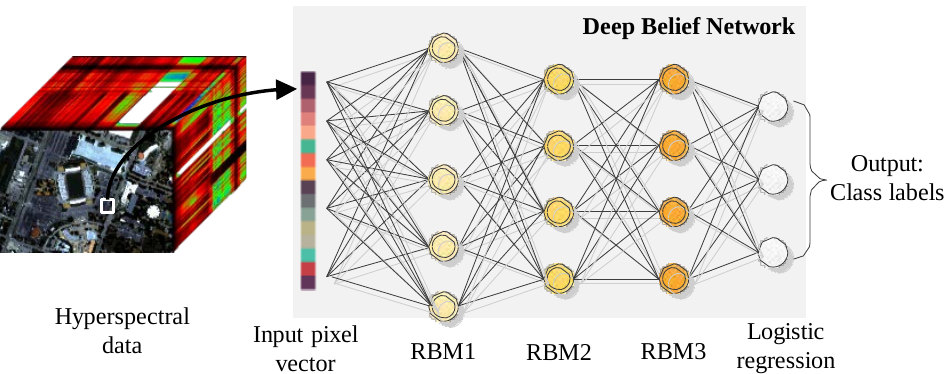}
 \caption{A spectral classifier based on DBN. The classification scheme shown here has four layers: one input layer, two RBMs, and one logistic regression layer \cite{Ghamisi-review-2017}.}
 \label{fig:DBN}
\end{figure}

In order to increase the feature representation capability of a single RBM, several
RBMs can be stacked one after another which establishes a DBN, which can be learned to extract a deep hierarchical representation of the training data. Fig. \ref{fig:DBN} illustrates a DBN composing of several RBM layers. In order to use DBN as a classifier, one can add a logistic regression layer at the end of the network to form a spectral classifier.

\subsection{CNNs}
\begin{figure*}
\begin{center}
\includegraphics[width=160mm]{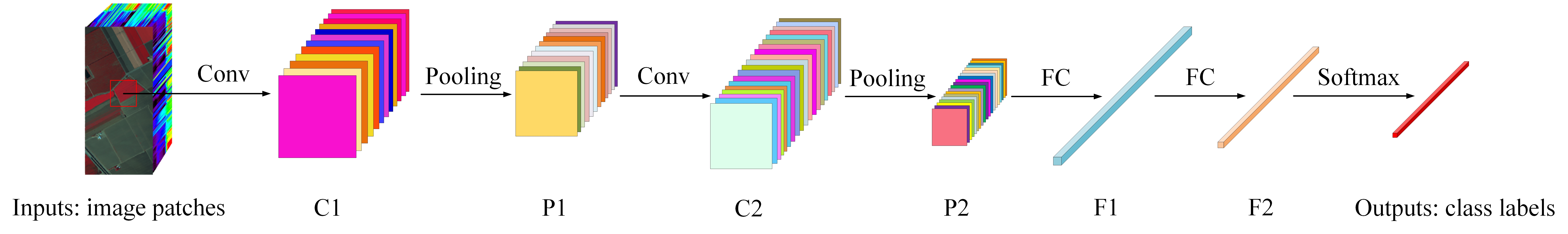}
\end{center}
\caption{The architecture of a conventional CNN, which consists of two convolutional-pooling layers and two fully connected layers.}
\label{CNN}
\end{figure*}

CNNs are inspired by the structure of the visual system. In contrast to fully connected networks, CNNs make use of local connections to extract the contextual 2-D spatial features of images. In addition, network parameters can be significantly reduced via the weight-share mechanism. A representative structure of CNNs is shown in Fig. \ref{CNN}, which mainly consists of a stack of alternating convolution layers and pooling layers with a number of fully connected layers. In the convolutional layers, the image patches with spatial context information are convolved with a set of kernels. Then, the pooling layers reduce the size of feature maps created by the convolutional layers to obtain more general and abstract features. Finally, these feature maps are further transformed into feature vectors via several fully connected layers. Here, each component is described as follows.

\subsubsection{Convolutional Layers}

In general, convolutional layers are the most important parts of CNNs. At each convolutional layer, the input cube is convoluted with multiple learnable filters, resulting in generating multiple feature maps. Specifically, let $\mathbf{X}$ be the input cube and its size is ${m}\times{n}\times{d}$, where ${m}\times{n}$ refers to the spatial size of $\mathbf{X}$, $d$ is the number of channels, and $\mathbf{x}_i$ is the $i$-th feature map of $\mathbf{X}$. Supposing there are $k$ filters at this convolutional layer and the $j$-th filter can be characterized by the weight $\mathbf{w}_j$ and \textit{bias} $b_j$. The $j$-th output of the convolutional layer can be represented as follows:

\begin{equation}
\mathbf{y}_j=\sum_{i=1}^d f(\mathbf{x}_i*{\mathbf{w}_j}+b_j), j=1,2,...,k
\end{equation}
where $*$ refers to the convolution operator and $f(\cdot)$ is an activation function which is utilized to improve the nonlinearity of the network. Recently, ReLU \cite{Alexnet} has been the mostly-used activation function. ReLU mainly has two advantages: a fast convergence and robustness for gradient vanishing \cite{gradient_vanishing}. The formulation of ReLU is denoted as:
\begin{equation}
\sigma(\mathbf{x})=max(0,\mathbf{x}).
\end{equation}
\par

\subsubsection{Pooling Layers}

Due to the existence of redundant information in images, the pooling layers are periodically inserted after several convolutional layers in the CNNs. With the pooling operation, the spatial size of the feature maps is progressively reduced, at the same time, the amount of parameters and computation of the network are also decreased. Through the pooling operations, the sizes of feature maps tend to shrink, and the representation of extracted features becomes more abstract. Specifically, for a $p\times{p}$ window-size neighbor denoted as $\mathbf{S}$, the average pooling operation can be denoted as:

\begin{equation}
z=\frac{1}{F} \sum_{(i,j)\in{\mathbf{S}}}x_{ij}
\end{equation}
where $F$ is the number of elements in $\mathbf{S}$ and $x_{ij}$ is the activation value corresponding to the position $(i,j)$.      \par

\subsubsection{Fully Connected Layers}

After the pooling layers, the feature maps of previous layer are flattened and fed to fully connected layers. In a traditional neural network, the fully connected layers are used to extract more deep and abstract features by reshaping feature maps into an n-dimension vector (e.g., n=4096 in AlexNet \cite{Alexnet}). In general, the fully connected layer can be defined as:
\begin{equation}
\mathbf{Y'}=\sum_{i=1}^C f(\mathbf{W}\mathbf{X'}+\mathbf{b})
\end{equation}
where $\mathbf{X'}$, $\mathbf{Y'}$, $\mathbf{W}$, and $\mathbf{b}$ refer to the input, output, weight, and \textit{bias} of a fully connected layer, respectively.

\subsection{RNNs}

RNNs were originally introduced in \cite{rnn_orig1, rnn_orig2}. In contrast to a feedforward neural network, an RNN can recognize patterns in sequences of data and dynamic temporal characteristics by using a recurrent hidden state whose activation at each step depends on that of the previous steps. \par
Let $\mathbf{x}=(\mathbf{x}_1,\mathbf{x}_2,\cdots,\mathbf{x}_T)$ be sequential data, where $\mathbf{x}_i$ is the data at the $i$-th time step. The recurrent hidden state $\mathbf{h}^{<t>}$ of the RNN at temporal sequence $t$ can be updated by
\begin{equation}\label{eq:bg1}
{\mathbf{h}^{<t>}=}
\begin{cases}
{0} & {\mbox{if $t=0$}}\\
{f_1(\mathbf{h}^{<t-1>},\mathbf{x}^{<t>})} & {\mbox{otherwise}}
\end{cases}\
\end{equation}
where $f_1$ is a nonlinear function (e.g., a logistic sigmoid function or hyperbolic tangent function). Conventionally, the update rule of the recurrent hidden state in (\ref{eq:bg1}) has been obtained by:
\begin{equation}\label{eq:bg2}
{\mathbf{h}^{<t>}=f_1(\mathbf{w}\mathbf{x}^{<t>}+\mathbf{u}\mathbf{h}^{<t-1>}+\mathbf{b}_h)}\
\end{equation}
where $\mathbf{w}$ and $\mathbf{u}$ represent the coefficient matrices for the input at the current step and for the activation of recurrent hidden units at the previous step, respectively, and $\mathbf{b}_h$ shows the corresponding bias vector. Then, $\mathbf{h}^{<t>}$ will be used to predict $\mathbf{y}^{<t>}$ at time step $t$ as follows:
 \begin{equation}\label{eq:bg21}
{\mathbf{y}^{<t>}=f_2(\mathbf{p}\mathbf{h}^{<t>}+\mathbf{b}_y)}\
\end{equation}
where $f_2$ is a nonlinear function, $\mathbf{p}$ is the coefficient matrix for the activation of recurrent hidden units at the current step, and $\mathbf{b}_y$ is the corresponding bias vector. As can be noticed, one can use different nonlinearities, $f_1$ and $f_2$, to estimate $\mathbf{h}^{<t>}$ and $\mathbf{y}^{<t>}$.

The performance of the conventional RNN can be downgraded when it deals with long-term sequential data due to the vanishing gradient or exploding gradient. To address this issue, long short-term memory (LSTM) \cite{lstm_1, lstm_2} and gated recurrent unit (GRU) \cite{chung2014empirical} were introduced.


As discussed above, RNNs are able to recognize patterns in sequences of data. In this context, RNNs are well-suited for the analysis of HSI since each pixel vector in hyperspectral data can be considered as a set of orderly and continuing spectra sequences in the spectral space. In \cite{Mou-RNN}, the concept of RNN was adopted for pixel-wise spectral classification by developing a new activation function and a modified gated recurrent unit, which can effectively analyze hyperspectral pixels as sequential data and then assign a classification label via network reasoning. 

\subsection{GANs}

In general, there are two typical classes of models in the machine learning community: generative approaches and discriminate approaches. Generative approaches try to learn the distribution parameters from data, and then they can generate new samples according to the learned models. Discriminate approaches attempt to model the dependence of labels (i.e., $\mathbf{y}$) on training data (i.e., $\mathbf{x}$), which can be used for predicting $\mathbf{y}$ from $\mathbf{x}$. \par

\begin{figure}
\begin{center}
\includegraphics[width=85mm]{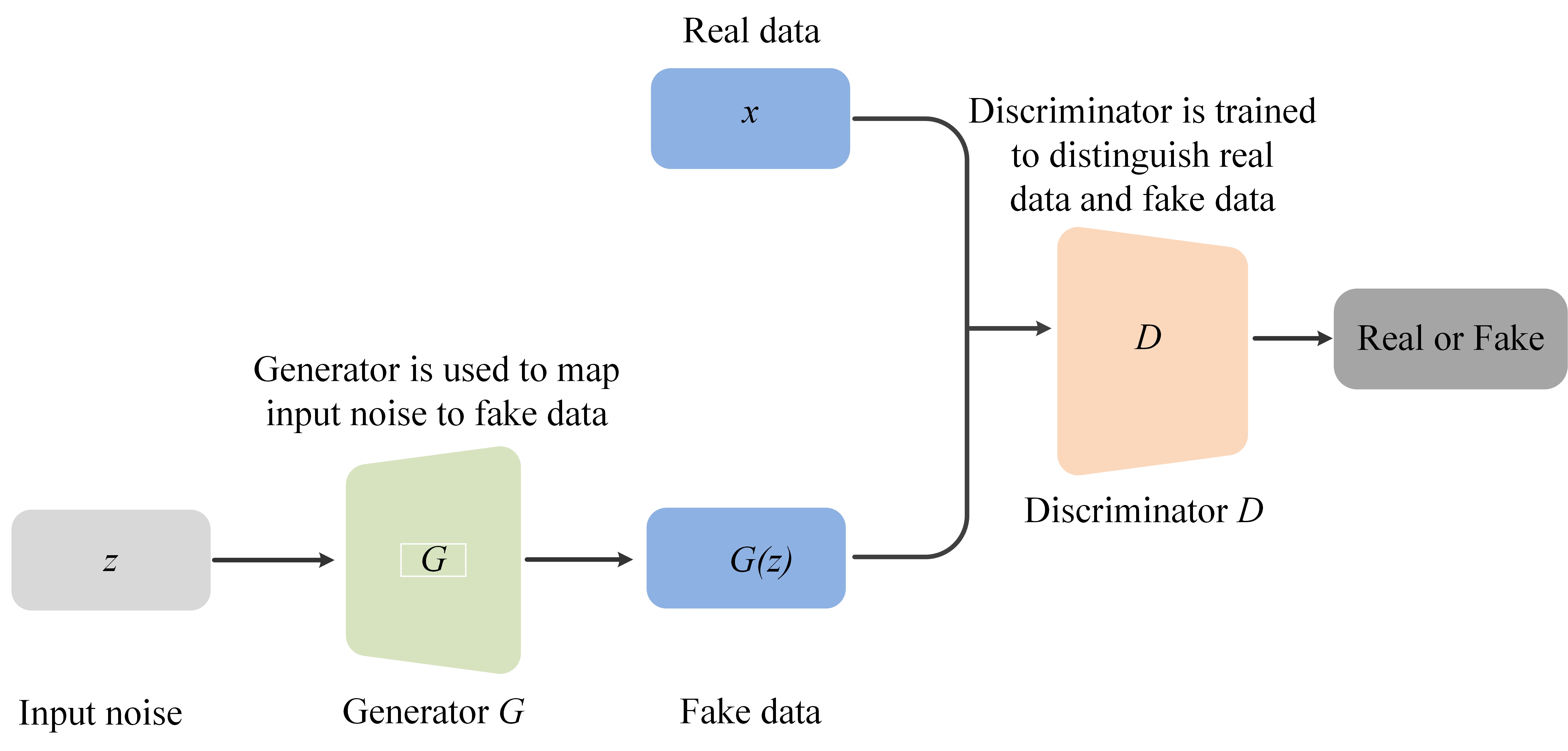}
\end{center}
\caption{A general illustration of a GAN \cite{Chen-GAN-HSI}.}
\label{GAN}
\end{figure}

GANs are the relatively new kind of models, which contains a generative model $\mathit{G}$ and a discriminative model $\mathit{D}$ \cite{GAN}. Models $\mathit{G}$ and $\mathit{D}$ are trained in an adversarial manner. Model $\mathit{G}$ tries to generate fake inputs as real as possible, whereas model $\mathit{D}$ tries to distinguish between real and fake inputs. Through the adversarial manner and competition of two models, the training process of the discriminator will proceed both continuously and effectively. The architecture of a GAN is shown in Fig. \ref{GAN}.

\begin{figure*}
\begin{center}
\includegraphics[width=160mm]{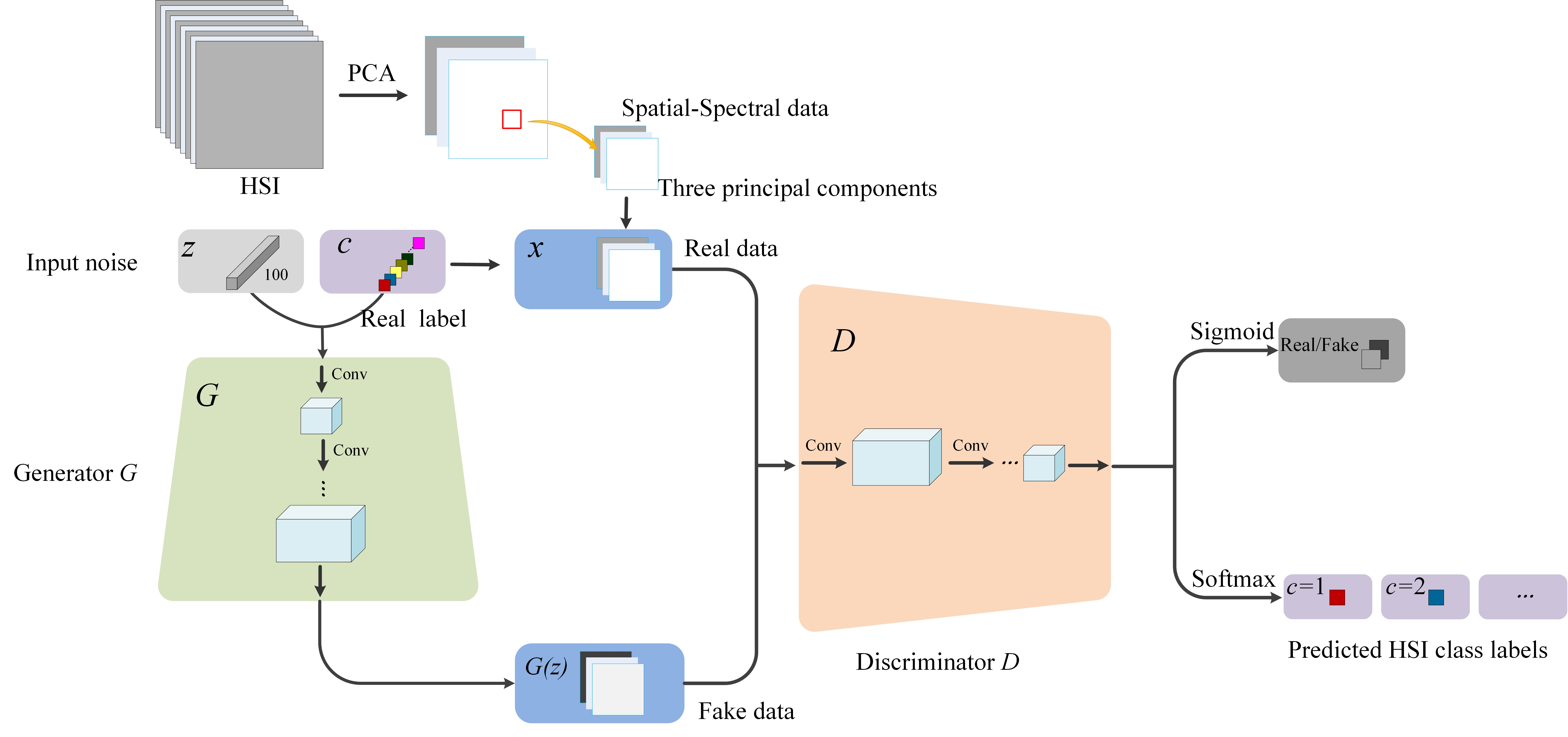}
\end{center}
\caption{Illustration of the GAN-based HSI classification approach developed in \cite{Chen-GAN-HSI}.}
\label{GAN-HSI}
\end{figure*}

The generator $\mathit{G}$ accepts random noise $\mathbf{z}$ as input and produces fake data $\mathit{G}(\mathbf{z})$. The discriminator $\mathit{D}$ estimates the probability that $\mathbf{x}$ is a true sample. The discriminator $\mathit{D}$ is trained to maximize $\log(\mathit{D}(\mathbf{x}))$, which is the probability of assigning the right labels to the training samples; the generator $\mathit{G}$ is trained to minimize $\log(1-\mathit{D}(\mathit{G}(\mathbf{z}))$. So, the aim of the GAN is to solve the following minimax problem

\begin{equation}
\begin{aligned}
\min \limits_\mathit{G} \max \limits_\mathit{D} \mathit{V}(\mathit{D},\mathit{G})= E_{\mathbf{x}\sim{p(\mathbf{x})}}\left[\log(\mathit{D}(\mathbf{x}))\right] \\
+E_{\mathbf{z}\sim{p(\mathbf{z})}}\left[\log(1-\mathit{D}(\mathit{G}(\mathbf{z})))\right]
\end{aligned}
\end{equation}
where $E$ is the expectation operator. Due to the advantages of CNN, the deep convolutional generative adversarial network architecture, which uses deep convolution networks in $\mathit{G}$ and $\mathit{D}$, was proposed in \cite{CRNN}. Although GAN is a promising technique, the original discriminator $\mathit{D}$ only estimates whether the input sample is either real or fake. Therefore, it is not suitable for multi-class classification. In \cite{Classifier-GANs}, Odena \emph{at al.} proposed an auxiliary classifier GAN, which can be used for classification, in which $\mathit{D}$ was modified to a softmax classifier that can output multi-class labels probabilities \cite{Classifier-GANs}.

A proposed framework of GAN-based HSI classification is shown in Fig. \ref{GAN-HSI}. From Fig. \ref{GAN-HSI}, we can see that the generator $\mathit{G}$ also accepts HSI class labels $c$ in addition to the noise $\mathbf{z}$ and the output of $\mathit{G}$ can be defined by $\mathbf{x}_{fake}=\mathit{G}(\mathbf{z})$. The training samples with corresponding class labels and the fake data generated by $\mathit{G}$ are regarded as the input of the discriminator $\mathit{D}$. The objective function contains two parts: the log-likelihood of the right source of input data $\mathit{L}_s$ and the log-likelihood of the right class labels $\mathit{L}_c$:

\begin{equation}
\begin{aligned}
\mathit{L}_s= E\left[\log\mathit{P}(\mathit{s}=real|\mathbf{x}_{real})\right] \\
+E\left[\log\mathit{P}(\mathit{s}=fake|\mathbf{x}_{fake})\right]
\end{aligned}
\end{equation}

\begin{equation}
\begin{aligned}
\mathit{L}_c= E\left[\log\mathit{P}(\mathit{c}=real|\mathbf{c}_{real})\right] \\
+E\left[\log\mathit{P}(\mathit{c}=fake|\mathbf{c}_{fake})\right].
\end{aligned}
\end{equation}
Therefore, $\mathit{D}$ is optimized to maximize the $\mathit{L}_s+\mathit{L}_c$ while $\mathit{G}$ is optimized to maximize the $\mathit{L}_c-\mathit{L}_s$.

\section{Deep Networks-based HSI Classification}

Recently, deep learning has become one of the most successful techniques and achieved impressive performance in the computer vision field \cite{DL}. Motivated by those great breakthroughs, deep learning has also been introduced to classify HSIs in the remote sensing field \cite{DL-RS-review1,DL-RS-review2,DL-RS-review3}. Compared with traditional hand-crafted features-based methods, deep learning can automatically learn high-level features from complex hyperspectral data. With these discriminative features, deep learning-based methods can effectively deal with the first problem mentioned above in Section I (i.e., the large spatial variability of spectral signature). Based on the fact, a large number of deep networks have been developed to extract features of HSIs and achieved good classification performance. However, the types of features extracted from deep networks may be different, e.g., spectral, spatial, and spectral-spatial features can be extracted by different deep networks. In this section, we systematically review the deep learning-based methods for HSI classification in a framework. In such framework, the deep networks used for HSI classification are divided into spectral-feature networks, spatial-feature networks, and spectral-spatial-feature networks. These different networks are expected to extract corresponding features for the subsequent classification. The following subsections will introduce the three kinds of networks in details.  \par

\subsection{Spectral-Feature Networks}

\begin{figure*}
\begin{center}
\includegraphics[width=160mm]{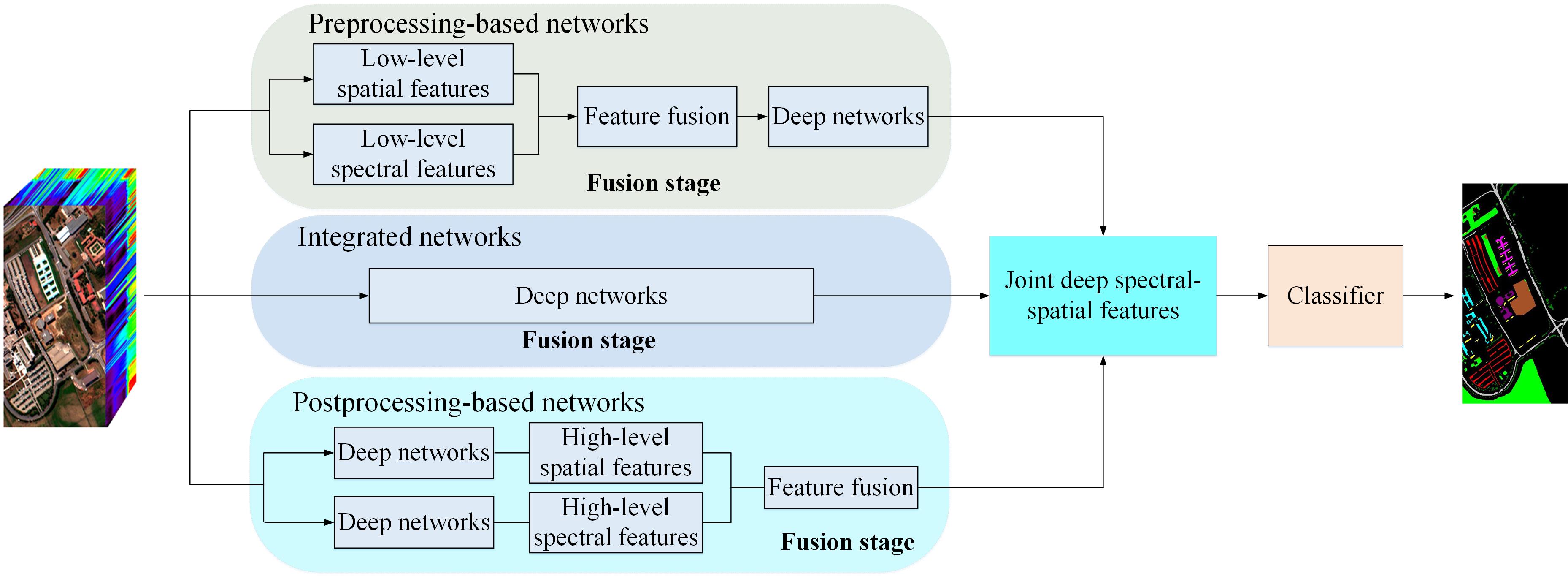}
\end{center}
\caption{Paradigms of spectral-spatial-feature networks, which can be further divided into three categories based on the stage of feature fusion, i.e., preprocessing-based networks, integrated networks, and postprocessing based networks.}
\label{SSFN}
\end{figure*}

Spectral information is the most important characteristic of HSIs and plays the vital role for the classification tasks \cite{Ghamisi-review-2017}. However, the hyperspecrtral remote sensors usually provide hundreds of spectral bands which also contain redundant information. Therefore, the direct exploration of original spectral vectors not only results in high computational cost but also decreases the classification performance. Although some traditional spectral feature extraction methods (e.g., PCA \cite{PCA1}, \cite{PCA2}, ICA \cite{ICA}, and LDA \cite{LDA}) can extract effective spectral features, the simple linear processing existed in these linear models is hard to handle the complex spectral property of HSIs. In this section, we introduce the deep learning-based framework, named spectral-feature networks, to extract deep spectral features. \par

In early research attempts, the hyperspectral pixel vector was intuitively fed into fully connected networks due to the requirement of the vector-based input in the network layers. Specifically, the original spectral vector was directly used to train the SAE or DBN in an unsupervised way \cite{Chen-SAE, Chen-DBN}. Subsequently, some improved methods were also developed for HSI classification. For example, Liu \emph{et al.} \cite{Active-DL} proposed an effective classification framework based on deep learning and active learning, where DBN was used to extract deep spectral features and an active learning algorithm was employed to iteratively select good-quality labeled samples as training samples. At the same time, Zhong \emph{et al.} \cite{Zhong-Diversity-DBN} developed an improved DBN model, named diversified DBN, to regularize the pretraining and fine-tuning procedures of DBN, which significantly improved the performance of DBN in terms of classification accuracies. In addition, 1-D CNN \cite{Hu-CNN-spectral, Chen-CNN, AL-B-CNN, HSIC_Deep_Models}, 1-D GAN \cite{Zhan-1-D-GAN, Chen-GAN-HSI}, and RNN \cite{Wu-1-D-CRNN, Mou-RNN, HSIC_Deep_Models} were also used to extract spectral features for HSI classification. In \cite{Li-CNN-DPPF}, Li \emph{et al.} used pixel-pair features extracted by CNN to explore correlation between hyperpsectral pixels, where the convolution operation was mainly executed in the spectral domain. Furthermore, in \cite{DDL-1, DDL-2}, the training of a deep network with the dictionary learning was reformulated. In such works, the multiple levels of dictionaries were learned in a robust fashion. This novel perspective obtained a better classification result than general deep networks.  \par

\subsection{Spatial-Feature Networks }

Previous researches on HSI classification have proven that classification accuracies can be further improved by incorporating spatial features into classifiers \cite{Pedram-HSIC-review,BenediktssonGhamisiBOOK}. In this section, we will discuss the spatial-feature networks that exploit deep networks to extract spatial features of HSIs. For accurate HSI classification, the learned spatial features are subsequently fused with spectral features extracted by other feature extraction techniques.  \par

In \cite{Chen-CNN, Yue-spatial-CNN, IGARSS2015-CNN-spatial, AL-B-CNN, SPDF}, PCA was first performed on the whole hyperspectral data to reduce the dimensionality of the original space, and then, the spatial information contained in the neighborhood region of the input hyperspectral pixel was exploited by a 2-D CNN. The above methods combined the PCA and CNN, which not only extracted the discriminative spatial features but also reduced the computational cost. More advanced, Liang \emph{et al.} \cite{Liang-SR-CNN} introduced a sparse representation technique to encode the deep spatial features extracted by a CNN into low-dimensional sparse features, which improved the ability of feature representation and the final classification accuracies. Chen \emph{et al.} adopted the off-the-shell CNNs (e.g., AlexNet \cite{Alexnet} and GoogLeNet \cite{googlenet}) to extracts deep spatial feature \cite{Cheng-Hierarchical}. In addition, Zhao \emph{et al.} \cite{BLDE-CNN} proposed a spectral-spatial feature-based classification (SSFC) framework for HSI classification. In that framework, the spectral and spatial features were extracted by the balanced local discriminant embedding (BLDE) and a CNN, respectively. Then, the spectral and spatial features were fused to train a multiple-feature-based classifier. Instead of extracting spatial-spectral information by incorporating the pixels within a small spatial neighborhood into a classifier, a novel HSI classification framework, named the deep multiscale spatial-spectral feature extraction algorithm, was proposed in \cite{Jiao-FCN-HSI}. In more detail, the well pretrained FCN-8 \cite{FCN} was first used to explore deep multiscale spatial structural information. Then, the weighted fusion mechanism was adopted to fuse the original spectral features and deep multiscale spatial features. Finally, the fused features were fed into a classifier to perform the classification operation.

\subsection{Spectral-Spatial-Feature Networks}

Instead of designing a deep network to extract either spectral features or spatial features, this kind of networks (i.e., spectral-spatial-feature networks) can extract joint deep spectral-spatial features for HSI classification. The joint deep spectral-spatial features are mainly obtained by the following three ways: 1) mapping the low-level spectral-spatial features to high-level spectral-spatial features via deep networks; 2) directly extracting deep features from original data or several principal components of the original data; 3) fusing two separate deep features (i.e., deep spectral features and deep spatial features). Based on this observation, the spectral-spatial-feature networks can be further divided into three categories: preprocessing-based networks, integrated networks, and postprocessing-based networks. Fig.~\ref{SSFN} depicts the paradigm of three kinds of networks, which is actually divided in terms of the processing stage, where the spectral information and the spatial information are fused. Here, the three networks are discussed in the following parts.   \par

\subsubsection{Preprocessing-based networks}

In this category, the spectral-spatial features are fused before feeding them into the subsequent deep network. In general, the whole classification process can be divided into three phases: 1) low-level spectral-spatial feature fusion; 2) high-level spectral-spatial feature extraction using deep networks; 3) joint deep spectral-spatial feature-based classification with simple classifiers (e.g., SVM, ELM, or multinomial logistic regression \cite{Ghamisi-review-2017}). As mentioned above, the fully connected networks (e.g., DBN, SAE, and their variants) only deal with 1-D inputs. To extract joint spectral-spatial features, a common idea is to flatten the spatial neighboring region into a 1-D vector, and then, the obtained spatial vector and the original spectral vector are stacked and fed into fully connected networks. For those works, readers can refer to \cite{Chen-SAE, Chen-DBN, Ma-SDAE, Grouped-DBN}. In \cite{Encoding-SAE, Hierarchical-CNN, Mei-sensor-specific}, a new spectral vector was first computed by averaging all spectral pixels within a spatial neighboring. Finally, the averaged spectral vector which actually includes the spatial contextual information was processed by the following deep network. Furthermore, instead of directly exploiting the spatial information within a neighboring window, some different filtering methods (e.g., Gobar filtering \cite{Kang-Gabor, Chen-Gabor}, attribute filtering \cite{AP-CNN}, extinction filtering \cite{EPs-CNN}, and rolling guidance filtering \cite{R-VCANet}) were introduced to process the original hyperspectral data aiming to extract more effective spatial features. These filter-based works combine the deep learning techniques with other spatial-feature extraction methods and deliver more accurate classification results.  \par

\subsubsection{Integrated networks}

\begin{figure}
\begin{center}
\includegraphics[width=80mm]{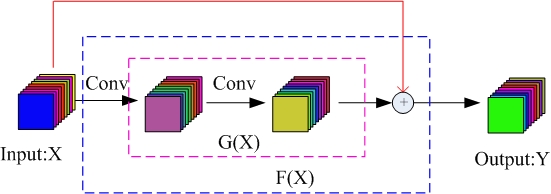}
\end{center}
\caption{Illustration of a residual block.}
\label{residual block}
\end{figure}

Instead of separately acquiring spectral and spatial features and then processing them together, the joint deep spectral-spatial features were directly extracted by the original data via 2-D CNN \cite{MCMs-2DCNN, Chen-HSI-LiDAR, Diverse-Region-CNN, CNN-MRF, DFFN, HDRN, Liu-simase-CNN, HSIC_Deep_Models, Deformable-CNN, Fang_SMBN}. Actually, the hyperspectral data can be typically represented in the format of a 3-D cube. Thus, 3-D convolution in spectral and spatial dimensions can naturally offer a more effective method for simultaneously extracting the spectral-spatial features within such images. Based on this fact, 3-D CNN was used to effectively extract deep spectral-spatial-combined features for accurate HSI classification without relying on any preprocessing or postprocessing techniques \cite{Chen-CNN, AL-B-CNN, Li-3-D-CNN, 3-D-DRN, 3-D-CNN-RS, HSIC_Deep_Models,3-D-CNN-GPU}. Apart from CNN, some powerful deep models were used to classify HSIs. Specifically, the fully convolutional network (FCN) \cite{FCN}, a very successful network in the semantic segmentation field, was utilized to reconstruct hyperspectral data, which can learn the deep features of HSIs in a supervised way \cite{FCN-HSIC} or an unsupervised way \cite{Mou-DRN-FCN}. Besides, references \cite{DFFN, Mou-DRN-FCN, going-deeper-HSI, 3-D-DRN, deep_pyramidal_residual} introduced residual learning \cite{DRN} to build very deep and wide networks with the purpose of extracting more discriminative features for HSI classification. Fig.~\ref{residual block} shows a residual block, where the output is the sum of input and convoluted value of input. In \cite{Chen-GAN-HSI, Wasserstein-GAN}, a 3-D GAN \cite{GAN} was utilized as a spectral-spatial classifier. In the GAN-based HSI classification framework, a CNN was first designed to discriminate the inputs, which is called the discriminative model. Then, another CNN was used to generate the so-called fake inputs as the generative model. Paoletti \emph{et al.} refined the CapsNets \cite{Capsule_Networks} as the spectral-spatial capsules to extract spectral-spatial features of HSIs \cite{Capsule_Networks_HSI}. Furthermore, researchers were also focused on building hybrid deep networks to achieve the accurate classification of HSIs, which can make full use of different deep models. For instance, Kemker \emph{et al.} \cite{self-taught} developed an unsupervised feature extraction framework to learn generalizable features from unlabeled hyperspectral pixels via a three-layer stacked convolutional autoencoder. Wu \emph{et al.} \cite{CRNN-pseudo-labels} proposed a novel deep convolutional recurrent neural networks (CRNN), taking advantages of CNN and RNN models, to classify HSIs by using pseudo labels. In such a network, the convolutional layers were used to extract middle-level locally invariant features from the input hyperspectral sequence, and the recurrent layers can extract contextual information from the middle-level feature sequences generated by the previous convolutional layers. In addition, a spectral-spatial cascaded RNN model was proposed to extract spectral-spatial features for HSI classification. \cite{Cascaded-RNN}. \par

\subsubsection{Postprocessing-based networks}

In this category, the whole classification procedure includes the following steps: 1) deep spectral features and deep spatial features are obtained via two deep networks; 2) the two kinds of features are fused in a fully connected layer to generate the joint deep spectral-spatial features; 3) joint deep spectral-spatial feature-based HSI classification with the subsequent classifier (e.g., SVM, ELM, or multinomial logistic regression \cite{Ghamisi-review-2017}). It is worth mentioning that the two deep networks used to separately extract deep spectral and spatial features may share the same weights or may be totaly different. For instance, in \cite{Learning-and-transferring, Multisource-RS}, a deep CNN with two-branch architecture was proposed to extract the joint spectral-spatial features from HSIs. In such works, a 1-D CNN branch and a 2-D CNN branch were used to extract spectral features and spatial features, respectively. After that, the learned spectral features and spatial features were concatenated and fed to fully connected layers to extract joint spectral-spatial features for classification. In \cite{Two-Stream}, Hao \emph{et al.} developed a novel two-stream architecture for HSI classification, where one stream employed a stacked denoising autoencoder (SDAE) to encode the spectral values of each input pixel, and the other stream exploited a deep CNN to learn spatial features by processing the corresponding image patch. Finally, the prediction probabilities from two streams were fused by adaptive class-specific weights, which can be obtained by a fully connected layer. In addition, references \cite{PCANet, Two-SSAE} used two similar network architectures (e.g., two principal component analysis-based networks \cite{PCANet}, two stacked sparse AE \cite{Two-SSAE}) to extract the spectral features and spatial features, respectively. Subsequently, these two kinds of features were fused in a fully connected layer and an SVM was further trained for classification. Furthermore, Santara \emph{et al.} \cite{BASS-Net} adopted multiple CNNs to simultaneously process multiple spectral sets, where multiple CNNs shared the same network parameters. In fact, each CNN aimed to extract the corresponding spatial feature of several neighboring spectral bands, and the spectral-spatial features can be obtained by fusing these individual spatial features in a fully connected layer.  \par

\section{Strategies for Limited Available Samples}

As a matter of fact, training a deep network requires a large number of training samples to learn the network parameters. However, in the remote sensing field, there is usually only a small amount of labeled data available since the collection of such labeled data is either expensive or time-demanding. This issue, also known as the imbalance between lots of weights and limited availability of training samples, may result in poor classification performance. Recently, some effective methods have been proposed to cope with the problem to some extent. In this section, we include some strategies to improve deep learning-based HSI classification under the condition of limited available samples.

\subsection{Data Augmentation}

Data augmentation is considered as an intuitional way to effectively solve the above problem. It tries to create new training samples from given known samples. By investigating the literature, we pointed out two main strategies to generate the additional virtual samples: 1) transformation-based sample generation; 2) mixture-based sample generation. In the following part, we will discuss the two methods in details.

\subsubsection{Transformation-based sample generation}

\begin{figure*}
\begin{center}
\includegraphics[width=140mm]{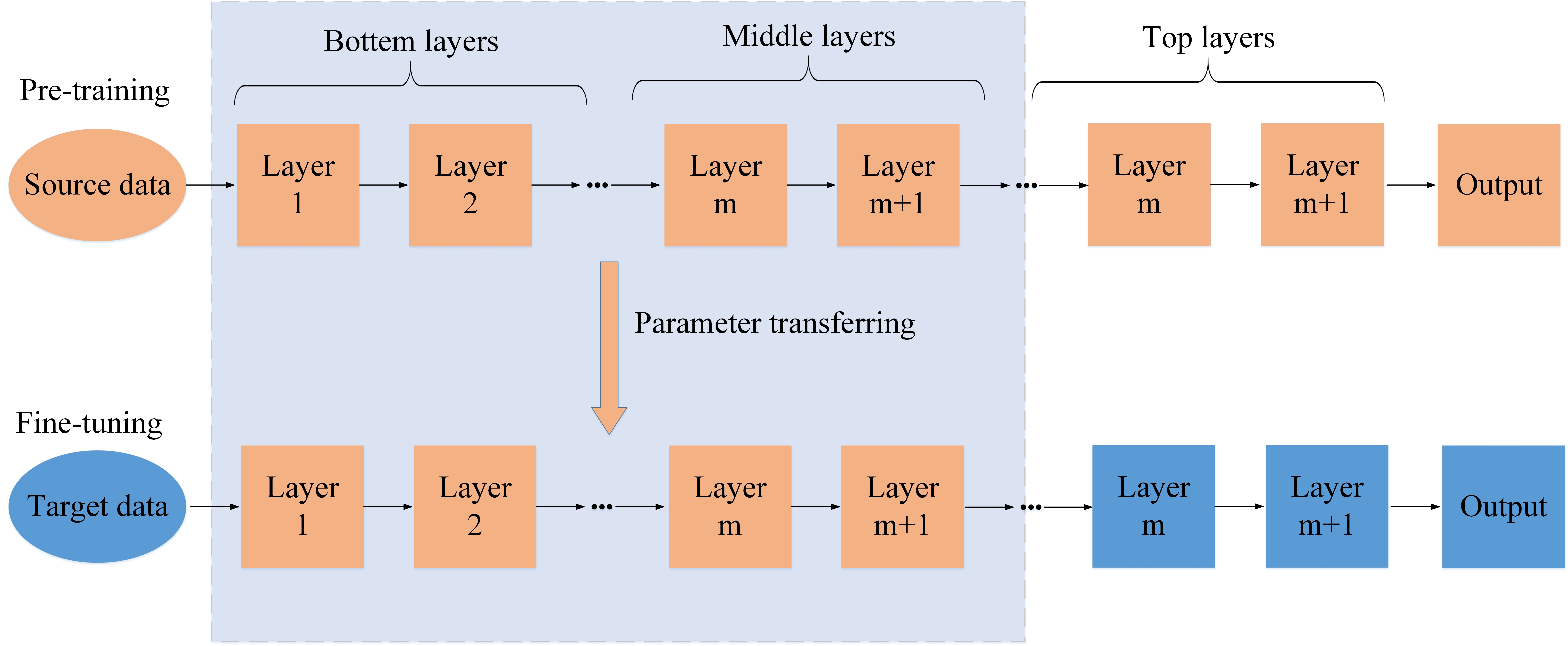}
\end{center}
\caption{The general framework of transfer learning based deep network training.}
\label{transfer learning framework}
\end{figure*}

Due to the complex situation of lighting in HSIs, objects of the same class in different locations may be affected by different radiations. Based on this fact, the virtual samples can be generated by transforming the current known samples. This method of augmenting data is widely used in \cite{Chen-CNN, Multisource-RS, going-deeper-HSI, AP-CNN, Diverse-Region-CNN}. Specifically, let $\mathbf{x}_i$ be a known training sample, the new virtual sample $\mathbf{y}$ can be obtained by
\begin{equation}
\mathbf{y}=f(\mathbf{x}_i)+\gamma{\mathbf{n}}
\end{equation}
where $f$ is a transforming function (e.g., performing the rotating, flipping, or mirroring operation), $\gamma$ controls the weight of the random Gaussian noise $\mathbf{n}$, which may be produced via the interaction of neighboring pixels or imaging error. Finally, the newly generated virtual sample $\mathbf{y}$ has the same class with $\mathbf{x}_i$ and can be used to train a deep network.

\subsubsection{Mixture-based sample generation}
In general, the objects of the same class usually show similar spectral characteristics in a certain range. This phenomenon makes it possible to generate a virtual sample from two given samples of the same class. In \cite{Chen-CNN, Kang-Gabor}, the virtual sample $\mathbf{y}$ is the linear combination of two training samples $\mathbf{x}_i$ and $\mathbf{x}_j$, which can be represented by the following formulation
\begin{equation}
{\mathbf{y}=\alpha_{ij}{\mathbf{x}_i}+(1-\alpha_{ij}){\mathbf{x}_j}}
\end{equation}
where $\alpha_{ij}$ represents the affinity between two training samples (i.e., $\mathbf{x}_i$ and $\mathbf{x}_j$) coming from the same class. $\alpha_{ij}$ is usually defined as follows:
\begin{equation}
\alpha_{ij}=exp(-\left\|\mathbf{x}_i-\mathbf{x}_j\right\|^2/2\sigma^2).
\end{equation}

Although there are many other techniques to generate the virtual samples, the above two methods are simple yet effective ways.

\subsection{Transfer Learning}

\begin{figure}
\begin{center}
\includegraphics[width=87mm]{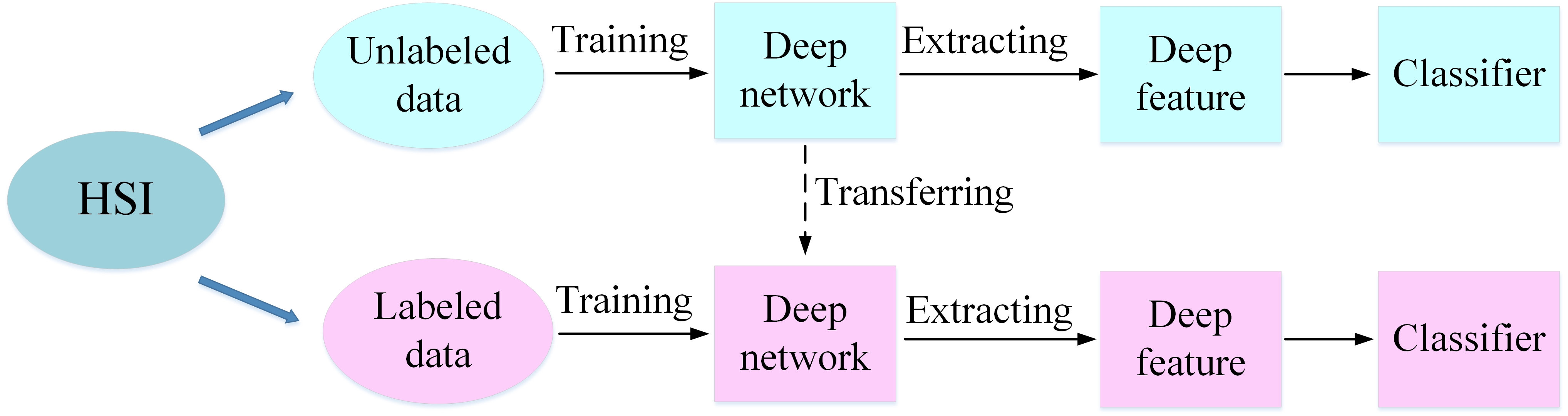}
\end{center}
\caption{Illustration of unsupervised/semisupervised feature learning associated with deep learning. The top part represents the unsupervised deep feature learning, and the bottom part illustrates the semisupervised way for feature learning. }
\label{feature-learning}
\end{figure}

Transfer learning \cite{transfer-learning} is regarded as a set of techniques that introduces the useful information learned from the source data to the target data, which can significantly decrease the demand on training samples. In recent years, transfer learning has been successfully applied in many fields, especially in the remote sensing field \cite{transferm-learning-in-RS-1, transferm-learning-in-RS-2} where the available training samples are not sufficient. In \cite{self-taught, Multisource-RS, Mei-sensor-specific, Learning-and-transferring, CRNN-pseudo-labels}, transfer learning was employed to set the initial values of network parameters copied from other trained deep networks, which provides better classification performance compared with the random initialization-based methods. Fig. \ref{transfer learning framework} shows the general framework of transfer learning for deep networks. Considering the fact that the low and middle layers usually capture generic features of input images which has high generalization for other images, it is feasible to directly transfer the network parameters of low and middle layers to a new network which has the same architecture as the previous network. Remarkably, the parameters of top layers are still initialized in a random way in order to deal with a specific task. Once the network parameters are transferred, the subsequent classification can be further divided into unsupervised and supervised methods. The former directly uses the deep features extracted from the transferred network to train the classifier. For the latter method, the network is further fine-tuned by using a small amount of training samples of the target data. In summary, by making full use of existing data sets, transfer learning can effectively solve the problem of degradation of network performance when the available training samples are limited.   \par

\subsection{Unsupervised/Semi-supervised Feature Learning}

Though supervised feature learning has gained great breakthrough in the HSI classification field, there is still an urgent need to learn HSI features in an unsupervised or semisupervised way. The main purpose of unsupervised/semisupervised feature learning is to extract useful features from an amount of unlabeled data. Recently, more and more research works \cite{Mou-DRN-FCN, self-taught, Recursive-AE, CRNN, Unsupervised-CNN, Two-SSAE, CRNN-pseudo-labels} focus on designing a robust and effective unsupervised/semisupervised feature learning framework based on deep learning to classify HSIs. Fig. \ref{feature-learning} illustrates unsupervised/semisupervised deep feature learning. From this figure, the top flowchart only uses unlabeled data to extract the informative features of HSIs, which is actually regarded as an unsupervised feature learning way. The deep network is elaborately designed as the encoder-decoder paradigm to learn networks without using label information. Furthermore, the classification performance can be improved via transferring the trained network and fine-tuning on the labeled data set, which is shown at the bottom of the flowchart in Fig. \ref{feature-learning}.    \par

For example, research works \cite{Chen-SAE, Chen-DBN, Zhong-Diversity-DBN, Two-SSAE, Active-DL, Ma-SDAE, self-taught, Recursive-AE} utilized a fully connected network to classify HSIs, where the training of the network was divided into pre-training in an unsupervised manner and fine-tuning with labeled data. In the pre-training stage, the sample from unlabeled data set was first mapped into an intermediate feature. Then, the intermediate feature was subsequently reconstructed via a decoding operation. At last, the parameters of each layer can be learned by minimizing the error between original sample and reconstructed sample. However, the above greedy layer-wise training fashion may not be efficient when the network is deep. Move advanced, an unsupervised end-to-end training framework was proposed in \cite{Mou-DRN-FCN}, where the convolutional and deconvolutional networks were regarded as encoder and decoder, respectively. In addition, the GAN \cite{GAN} was also adopted to construct semisupervised feature learning framework for HSI classification \cite{Zhan-1-D-GAN, Chen-GAN-HSI}. In such works, the generator created fake hyperspectral samples that were similar to the real data to train the GAN. Besides, Wu \emph{et al.} used abundant unlabeled data with their pseudo labels obtained by a non-parametric Bayesian clustering algorithm to pre-train a CRNN for HSI classification \cite{CRNN-pseudo-labels}. The semisupervised learning mechanism can extract more representative spectral-spatial features and provide satisfactory classification result.  \par

\subsection{Network Optimization}

The main purpose of network optimization is to further improve the network performance through adopting more efficient modules or functions. For instance, the current mainstream deep networks use ReLU \cite{Alexnet} as nonlinear activation function followed by the batch normalization operation \cite{Batch-Normalization}, which can effectively alleviate the overfitting during the training phase. In addition, research works \cite{DFFN, HDRN, 3-D-DRN, Mou-DRN-FCN, going-deeper-HSI} introduce residual learning \cite{DRN} to build a very deep network with the purpose of extracting more discriminative features for HSI classification. By optimizing several convolutional layers as the identity mapping, residual learning can ease the training process. Fig. \ref{residual block} shows the mechanism of residual learning. Let $ \mathbf{X} $ be the input of the first layer and $ F(\mathbf{X}) $ be the original underlying function to be learned by two stacked convolutional layers. The residual learning actually attempts to optimize the two convolutional layers as identity mapping, which is achieved by using skip connections (the red line in Fig. \ref{residual block}). By introducing residual function $G(\mathbf{X})=F(\mathbf{X})-\mathbf{X}$, the objective function from the original $F(\mathbf{X})=\mathbf{X}$ equivalently converts to $G(\mathbf{X})=0$. As described above, $F(\mathbf{X})$ can be written by the following formulation:
\begin{equation}
       F(\mathbf{X})=G(\mathbf{X})+\mathbf{X}.
\end{equation}
As shown in Fig. \ref{residual block}, $G(\mathbf{X})$ can be computed by two convolutions with $\mathbf{X}$:
\begin{equation}\label{Eq2}
       G(\mathbf{X})=f(f(\mathbf{X}*\mathbf{W}_1+\mathbf{b}_1)*\mathbf{W}_2+\mathbf{b}_2)
\end{equation}
where $\mathbf{W}_1$ and $\mathbf{W}_2$ are convolutional kernels, $\mathbf{b}_1$ and $\mathbf{b}_2$ are trainable \textit{bias} parameters, and $f$ refers to the ReLU \cite{Alexnet} function. By stacking multiple residual blocks, the extracted features become more and more discriminative and the accuracy is gained from considerably increased depth. More details about theoretical and experimental proofs of the DRN can refer to \cite{DRN}.  \par

\begin{figure}
\begin{center}
\subfigure[]{\includegraphics[width=90mm]{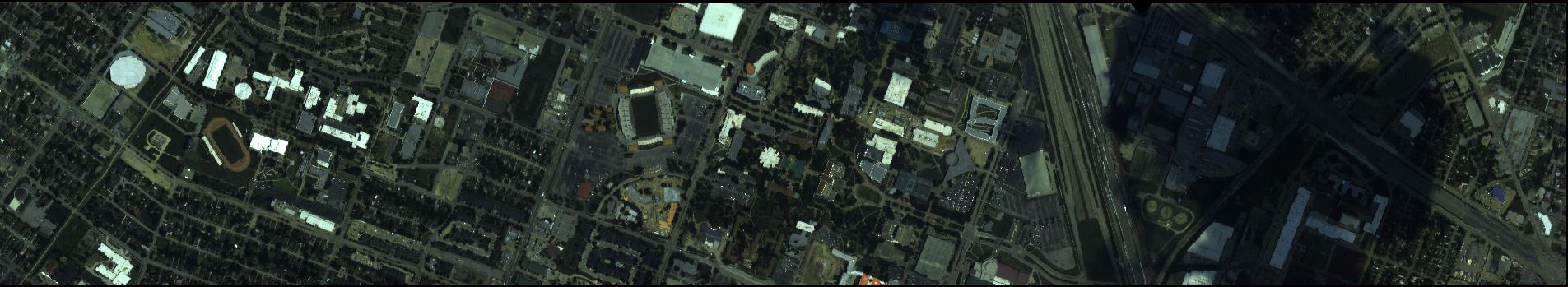}}
\subfigure[]{\includegraphics[width=90mm]{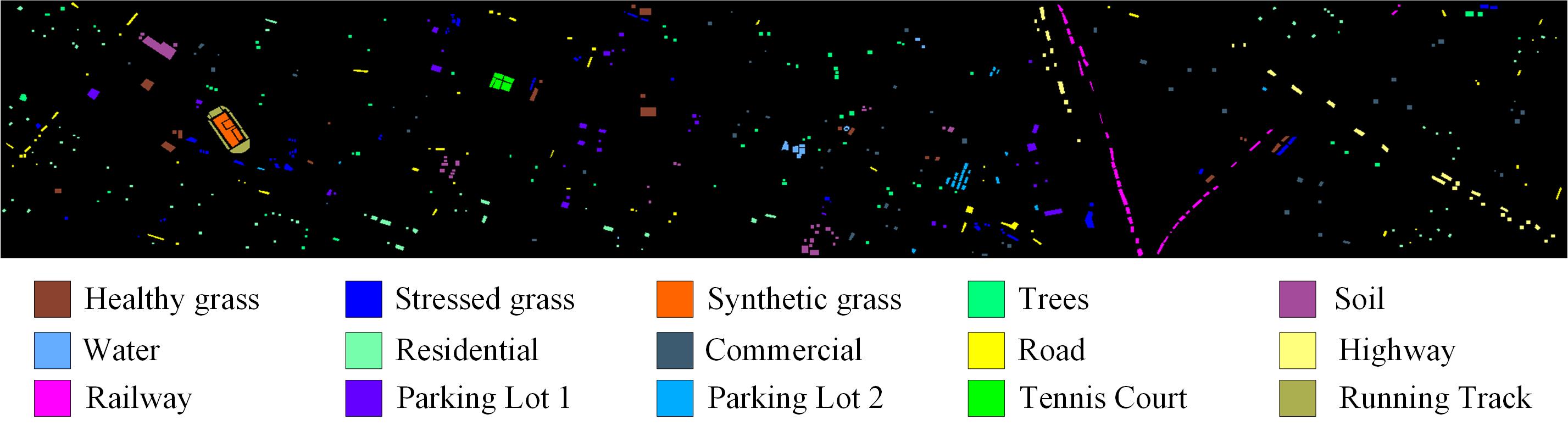}}
\caption{The Houston data set. (a) The three-band false color composite. (b) Ground reference data and color code. }
\label{Houston_fig}
\end{center}
\end{figure}

Besides, other network optimization strategies are also exploited to boost the representation ability of features extracted by deep networks. For example, Zhong \emph{et al.} \cite{Zhong-Diversity-DBN} improved the performance of DBN via regularizing the pre-training and fine-tuning procedure, which delivers a better classification accuracy than usual DBN models. In \cite{Encoding-SAE}, the label consistency constraint was enforced into the training procedure of SAE. Moreover, the correlation between samples was considered in networks investigated in \cite{Li-CNN-DPPF, Liu-simase-CNN}.  \par

In general, network optimization is still a challenging problem. In the future works, the unique characteristics of HSI should be considered as much as possible when designing an improved network for HSI classification.   \par

\begin{figure}
\begin{center}
\subfigure[]{\includegraphics[width=25mm, height=50mm]{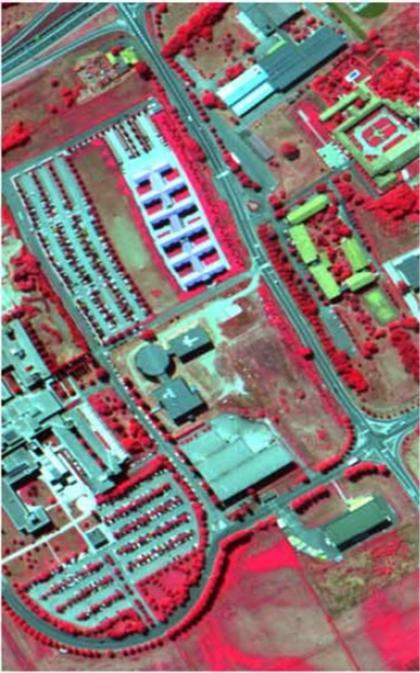}}
\subfigure[]{\includegraphics[width=25mm, height=50mm]{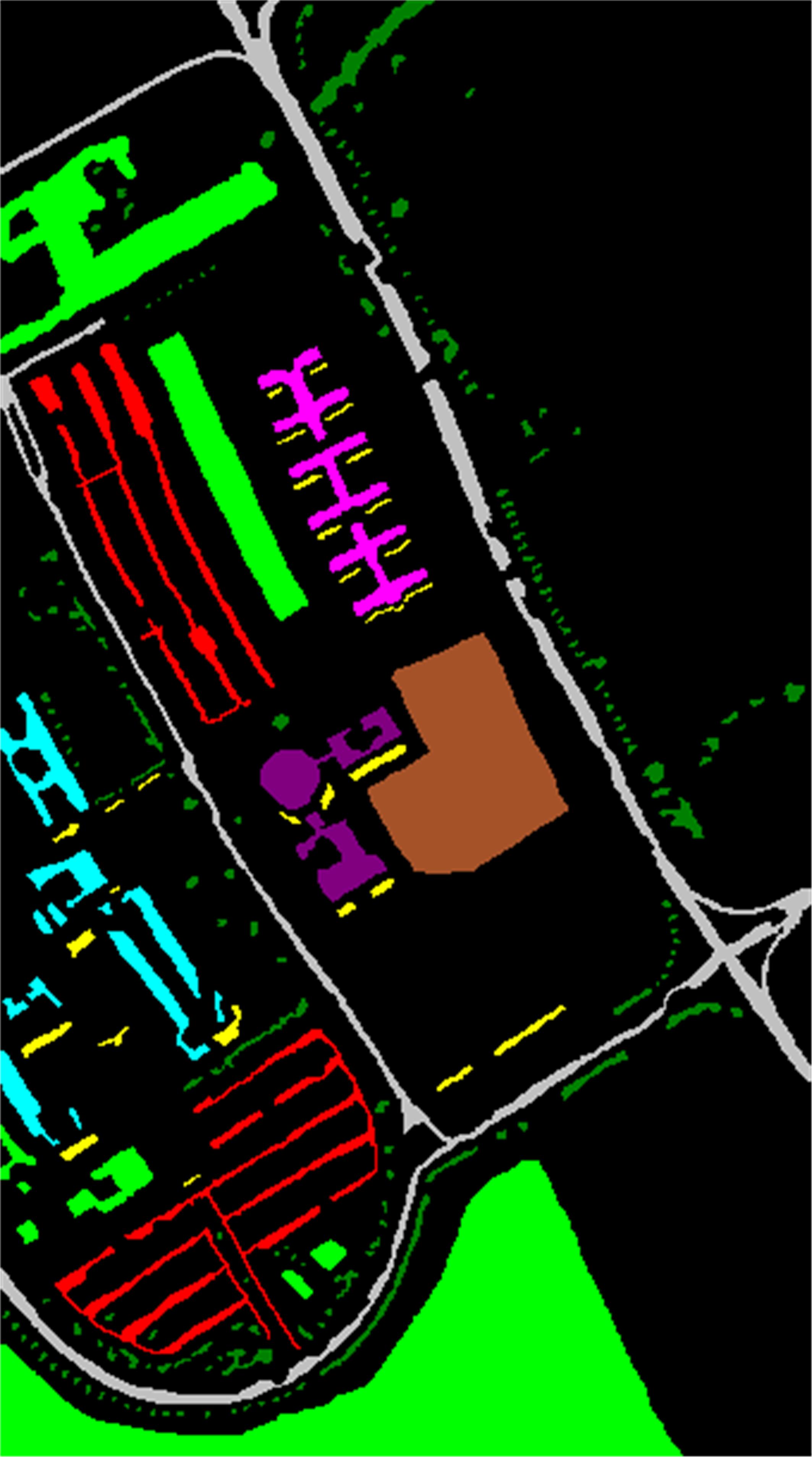}}
            {\includegraphics[width=15mm]{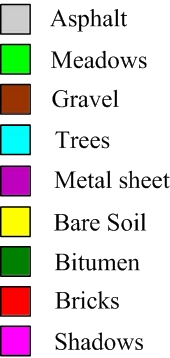}}
\caption{The University of Pavia data set. (a) The three-band false color composite. (b) Ground reference data and color code. }
\label{PU_fig}
\end{center}
\end{figure}

\section{Experiments}

\begin{figure}
\begin{center}
\subfigure[]{\includegraphics[width=25mm, height=50mm]{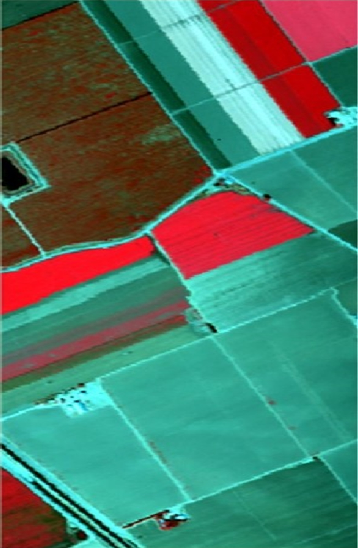}}
\subfigure[]{\includegraphics[width=25mm, height=50mm]{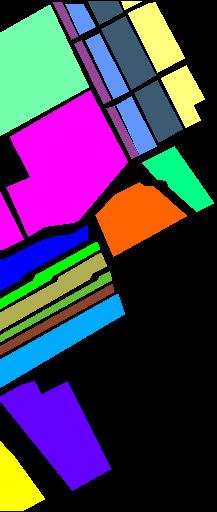}}
            {\includegraphics[width=19mm]{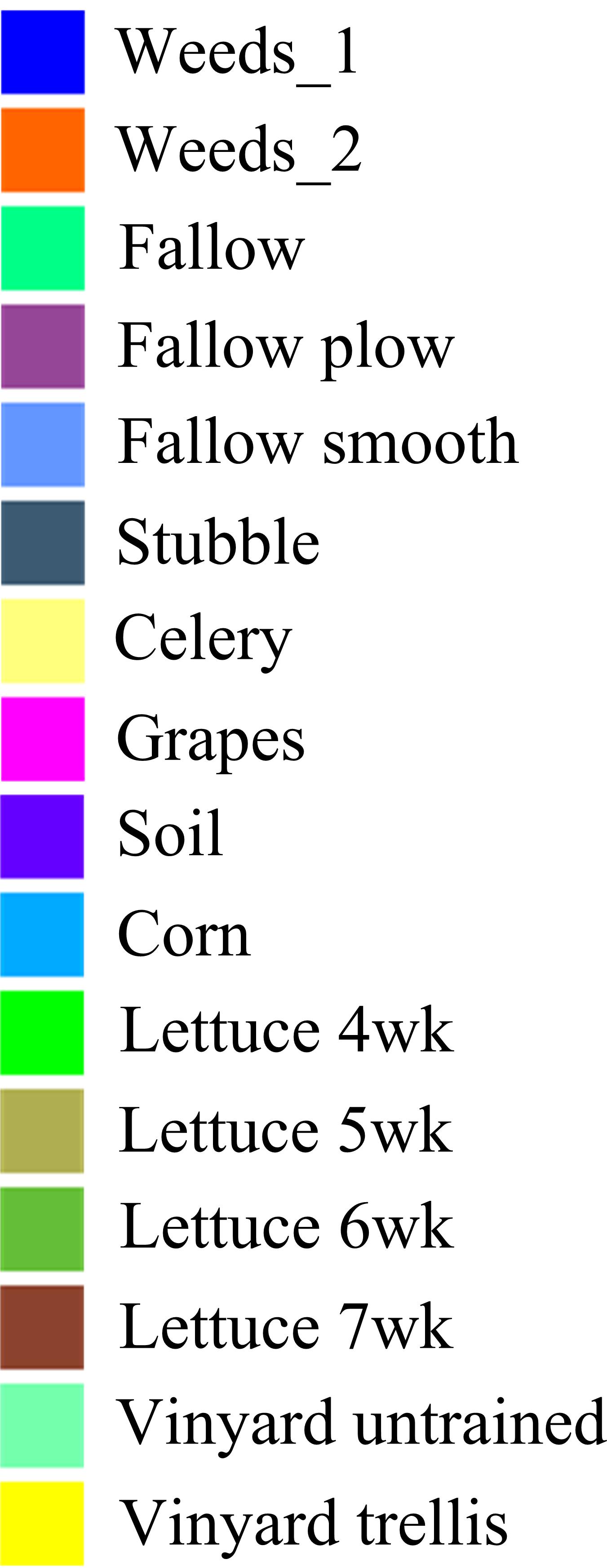}}
\caption{The Salinas data set. (a) The three-band false color composite. (b) Ground reference data and color code. }
\label{Salinas_fig}
\end{center}
\end{figure}

\begin{figure}[tp]
\centering
\subfigure[SVM]{\includegraphics[width=78mm]{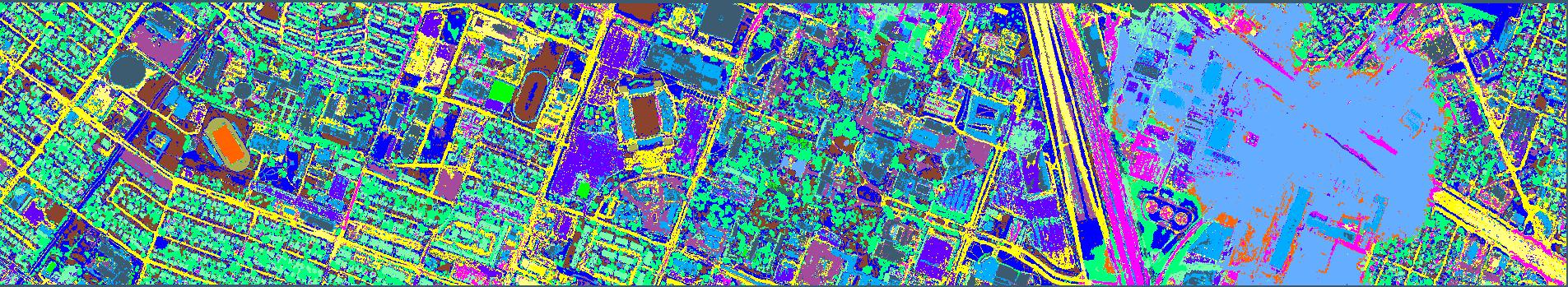}}
\hspace{+1pt}
\subfigure[EMP]{\includegraphics[width=78mm]{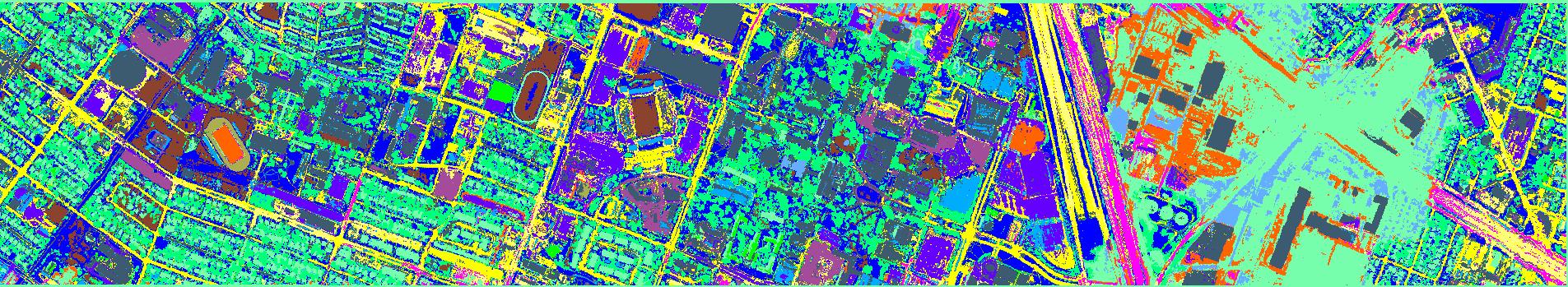}}
\hspace{+1pt}
\subfigure[JSR]{\includegraphics[width=78mm]{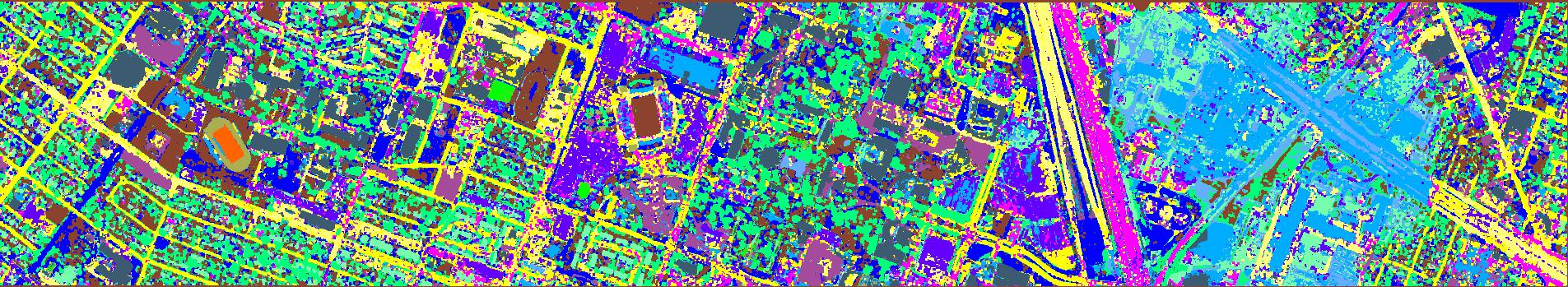}}
\hspace{+1pt}
\subfigure[EPF]{\includegraphics[width=78mm]{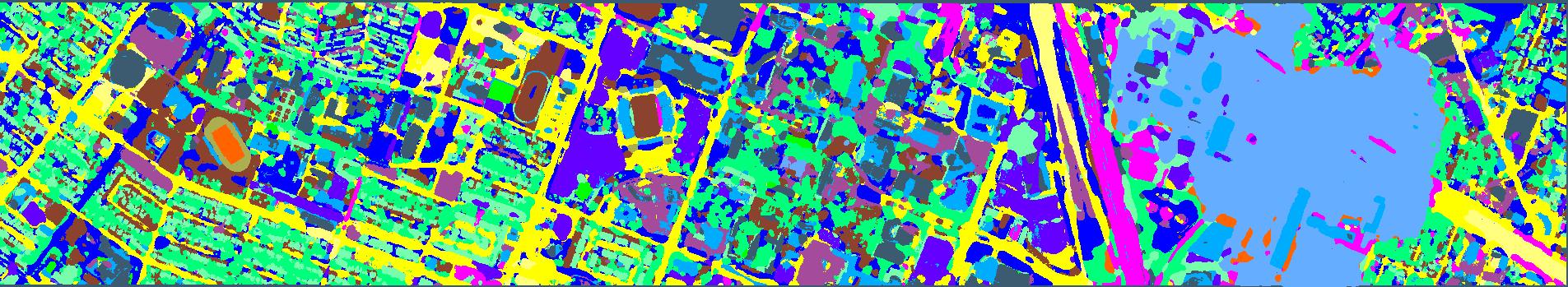}}
\hspace{+1pt}
\subfigure[3D-CNN]{\includegraphics[width=78mm]{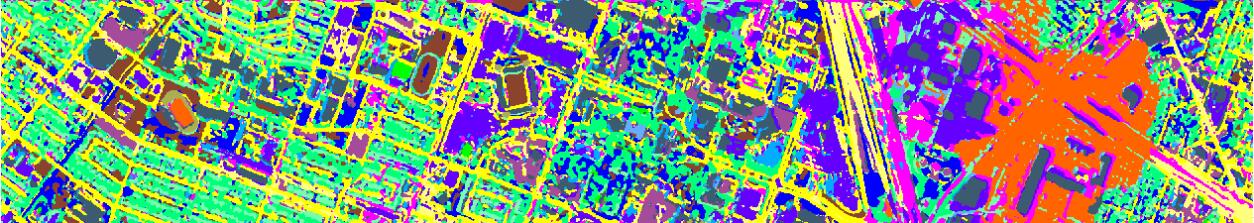}}
\hspace{+1pt}
\subfigure[CNN-PPF]{\includegraphics[width=78mm]{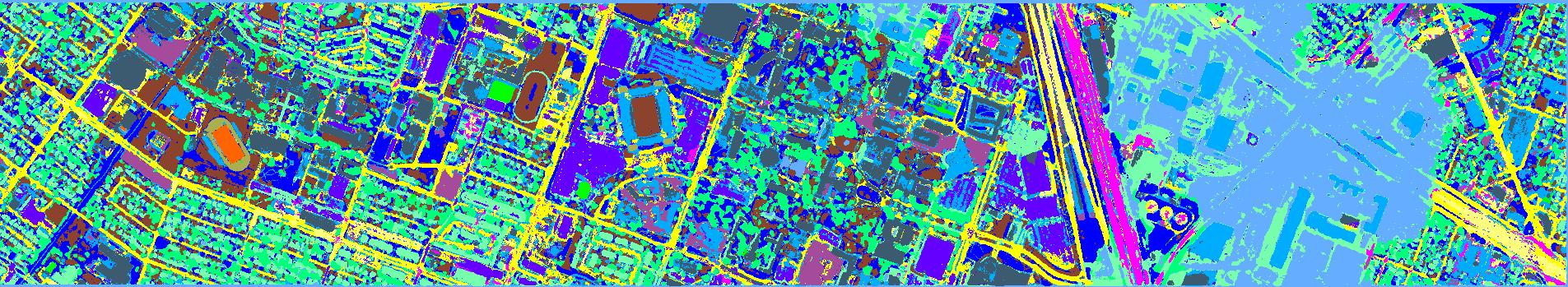}}
\hspace{+1pt}
\subfigure[Gabor-CNN]{\includegraphics[width=78mm]{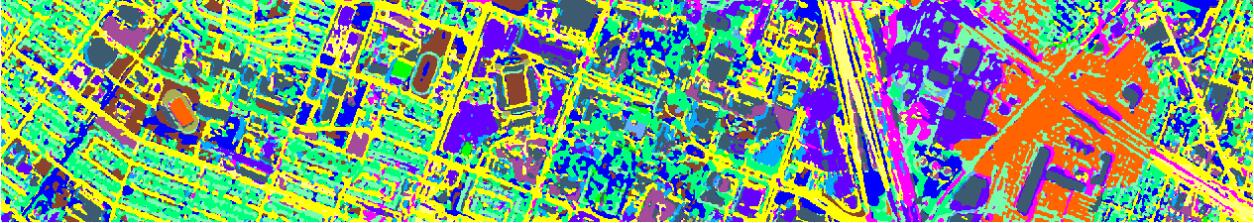}}
\hspace{+1pt}
\subfigure[S-CNN]{\includegraphics[width=78mm]{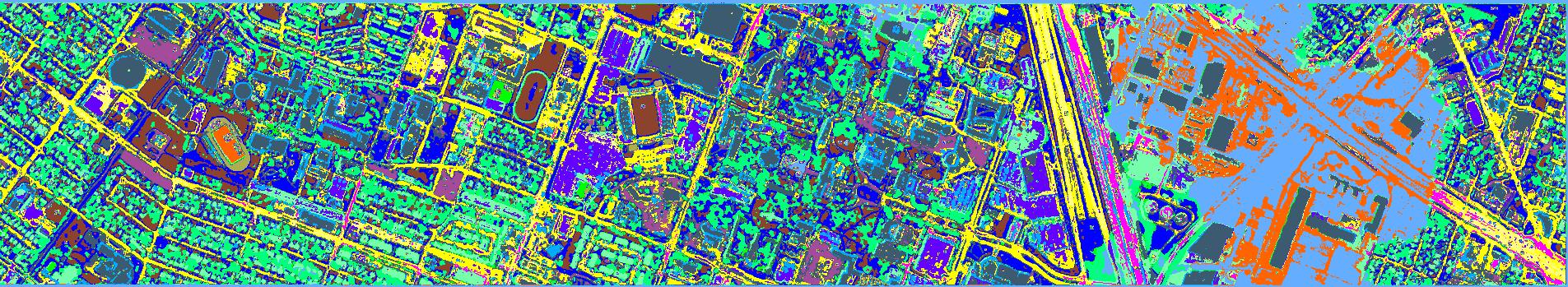}}
\hspace{+1pt}
\subfigure[3D-GAN]{\includegraphics[width=78mm]{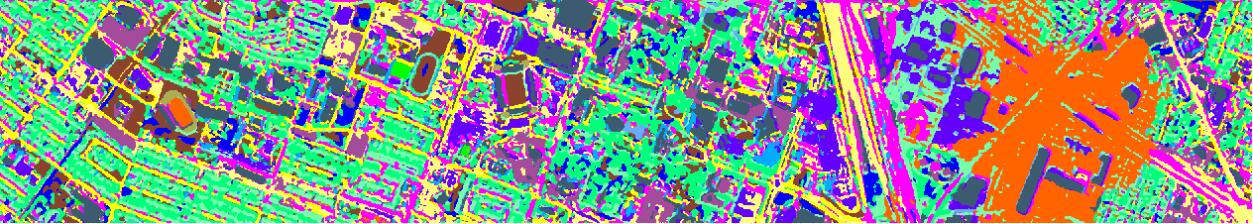}}
\hspace{+1pt}
\subfigure[DFFN]{\includegraphics[width=78mm]{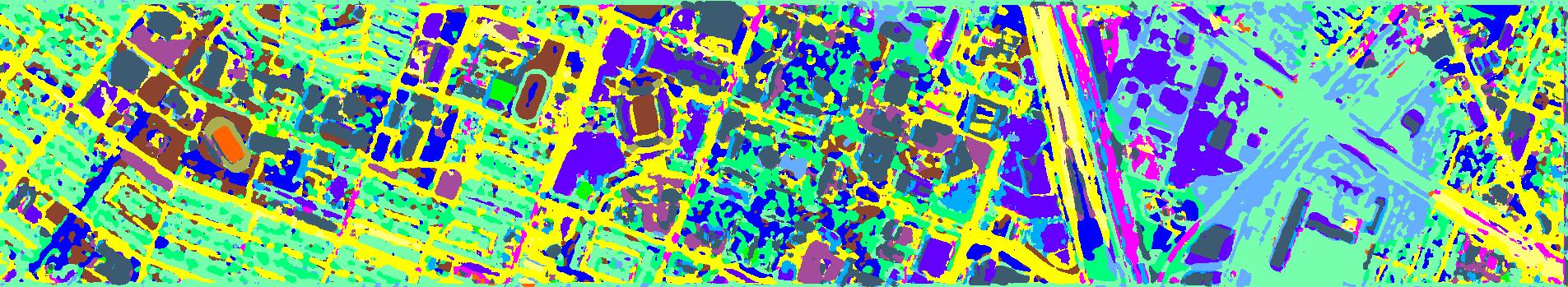}}

\caption{Classification maps for the Houston data set obtained by (a) SVM \cite{SVM}, (b) EMP \cite{EMP1}, (c) JSR \cite{JSR}, (d)  EPF \cite{EPF}, (e) 3D-CNN \cite{Chen-CNN}, (f) CNN-PPF \cite{Li-CNN-DPPF}, (g) Gabor-CNN \cite{Chen-Gabor}, (h) S-CNN \cite{Liu-simase-CNN}, (i) 3D-GAN \cite{Chen-GAN-HSI}, and (j) DFFN \cite{DFFN}.}
\label{Houston_classification_maps}
\end{figure}

\begin{table}
\centering
\footnotesize
  \caption{Number of Training and Test Samples Used for the Houston Data}
  \label{Houston_samples}
\begin{tabular}{ccccc}
\hline
\hline
 \multirow{2}{*} {Class name}          & Training             & Test      & Total\\
               & samples              & samples   & samples\\
\hline
 Healthy grass          & 198                   & 1053    & 1251 \\
 Stressed grass         & 190                   & 1064    & 1254 \\
 Synthetic grass        & 192                   & 505     & 697 \\
 Trees                  & 188                   & 1056    & 1244 \\
 Soil                   & 186                   & 1056    & 1242 \\
 Water                  & 182                   & 143     & 325 \\
 Residential            & 196                   & 1072    & 1268 \\
 Commercial             & 191                   & 1053    & 1244 \\
 Road                   & 193                   & 1059    & 1252 \\
Highway                & 191                   & 1036     & 1227 \\
 Railway                & 181                   & 1054    & 1235 \\
 Parking Lot 1          & 192                   & 1042    & 1234\\
 Parking Lot 2          & 184                   & 285     & 469 \\
 Tennis Court           & 181                   & 247     & 428 \\
 Running Track          & 187                   & 473     & 660 \\
\hline
 Total                  & 2832                  & 12179   & 15011 \\
\hline
\hline
\end{tabular}
\end{table}

\begin{table}
\centering
\footnotesize
  \caption{Number of Training and Test Samples Used for the University of Pavia Data Set}
  \label{PU_samples}
\begin{tabular}{cccc}
\hline
\hline
   \multirow {2}{*}{Class name}          & Training             & Test      & Total\\
               & samples              & samples   & samples\\
\hline
 Asphalt                & 200                   & 6431     & 6631  \\
 Meadows                & 200                   & 18449    & 18649 \\
 Gravel                 & 200                   & 1899     & 2099 \\
 Trees                  & 200                   & 2864     & 3064 \\
 Painted metal sheets   & 200                   & 1145     & 1347 \\
 Bare Soil              & 200                   & 4829     & 5029 \\
 Bitumen                & 200                   & 1130     & 1330 \\
 Self-Blocking Bricks   & 200                   & 3482     & 3682 \\
 Shadows                & 200                   & 747      & 947 \\
\hline
 Total                  & 1800                  & 40976    & 42776\\
\hline
\hline
\end{tabular}
\end{table}

\begin{table}
\centering
\footnotesize
  \caption{Number of Training and Test Samples Used for the Salinas Data Set}
  \label{Salinas_samples}
\begin{tabular}{cccc}
\hline
\hline
  \multirow{2}{*}{Class name}          & Training             & Test      & Total\\
           & samples              & samples   & samples\\
\hline
 Brocoli\_green\_weeds\_1  & 200                   & 1809   & 2009 \\
 Brocoli\_green\_weeds\_2  & 200                   & 3526   & 3726 \\
 Fallow                 & 200                   & 1776      & 1976 \\
 Fallow\_rough\_plow      & 200                   & 1194    & 1394 \\
 Fallow\_smooth          & 200                   & 2478     & 2678 \\
 Stubble                & 200                   & 3759      & 3959 \\
 Celery                 & 200                   & 3379      & 3579 \\
 Grapes\_untrained       & 200                   & 11071    & 11271 \\
 Soil\_vinyard\_develop   & 200                   & 6003    & 6203 \\
 Corn\_senesced\_green\_weeds & 200               & 3078    & 3278 \\
 Lettuce\_romaine\_4wk    & 200                   & 868     & 1068 \\
 Lettuce\_romaine\_5wk    & 200                   & 1727    & 1927 \\
 Lettuce\_romaine\_6wk    & 200                   & 716     & 916 \\
 Lettuce\_romaine\_7wk    & 200                   & 870     & 1070 \\
 Vinyard\_untrained      & 200                   & 7068     & 7268 \\
 Vinyard\_vertical\_trellis & 200                 & 1607    & 1807  \\
\hline

            Total                  & 3200                  & 50929     & 54129 \\
\hline
\hline
\end{tabular}
\end{table}

In this section, we mainly conduct a comprehensive set of experiments from four aspects. Firstly, a series of experiments are designed to demonstrate the advantages of deep learning on HSI classification over traditional methods. Secondly, the classification performance of several recent state-of-the-art deep learning approaches is systematacially compared. Thirdly, we visualize the learned deep features and network weights to further explore the ``black box". Finally, the effectiveness of strategies included in Section IV is further analyzed. To complete our experiments, three benchmark HSIs are used, i.e., the Houston, University of Pavia, and Salinas images. The three images are introduced in the following subsection.  \par

\subsection{Experimental Data Sets}

The Houston data was distributed for the 2013 IEEE Geoscience and Remote Sensing Society (GRSS) data fusion contest. This scene was captured in 2012 by an airborne sensor over the area of University of Houston campus and the neighboring urban area. The size of the data is $349\times{1905}$ pixels with a spatial resolution of 2.5 m. This HSI consists of 144 spectral bands with wavelength ranging from 0.38 to 1.05 $\mu$m and includes 15 classes. Fig.~\ref{Houston_fig} shows the false color composite of the Houston data set and the corresponding ground reference data, respectively. Table \ref{Houston_samples} shows the information on the number of training and test samples for the different classes of interests.  \par

The University of Pavia data set, which captures an urban area surrounding the University of Pavia, Italy, was collected by the ROSIS-03 sensor in Northern Italy in 2001. This scene is of size $610\times{340}\times{115}$ with a spatial resolution of 1.3 m per pixel and spectral coverage ranging from 0.43 to 0.86 $\mu$m. This image includes 9 classes of interest and has 103 spectral bands after removing 12 very noisy bands. Fig.~\ref{PU_fig} shows the false color composite of the University of Pavia image and the corresponding ground reference map, respectively. Table \ref{PU_samples} shows information on the number of training and test samples for the different classes of interests. \par

The Salinas data set was captured by the AVIRIS sensor over Salinas Valley, California. This image comprises of $512\times{217}$ pixels with a spatial resolution of 3.7 m and 204 bands after removing 20 water absorption bands. The available ground reference map covers 16 classes of interest. The false color composite of Salinas image and the corresponding ground reference data are shown in Fig.~\ref{Salinas_fig}. Table \ref{Salinas_samples} shows information on the number of training and test samples for the different classes of interests. \par

\begin{table*}
\centering
\footnotesize
  \caption{The Classification Accuracies (in Percentages) Obtained by SVM \cite{SVM}, EMP \cite{EMP1}, EPF \cite{EPF},  JSR \cite{JSR}, 3D-CNN \cite{Chen-CNN}, Gabor-CNN \cite{Chen-Gabor}, CNN-PPF \cite{Li-CNN-DPPF}, S-CNN \cite{Liu-simase-CNN}, 3D-GAN \cite{Chen-GAN-HSI}, and DFFN \cite{DFFN} on the Houston Image. The Best Accuracies Are Marked in Bold.}
\begin{tabular}{ccccccccccc}
\hline
\hline
classes	&	SVM	&	EMP	&	JSR	&	EPF	&	3D-CNN	&	CNN-PPF	&	Gabor-CNN	&	S-CNN	&	3D-GAN	&	DFFN \\
\hline
1	&	82.05	&	81.32	&	\textbf{94.49}	&	82.72	&	82.16	&	82.66	&	82.30	& 	83.00	&	81.58	&	77.41 \\
2	&	80.55	&	82.65	&	83.46	&	82.71	&	\textbf{85.19}	&	84.02	&	84.24	& 	83.27	&	79.74	&	81.39 \\
3	&	\textbf{100}	&	98.67	&	98.42	&	99.78	&	90.69	&	\textbf{100}	&	95.61	& 	98.66	&	97.42	&	94.59 \\
4	&	92.52	&	84.26	&	\textbf{96.59}	&	93.09	&	93.35	&	92.87	&	92.39	& 	93.50	&	93.36	&	87.82 \\
5	&	98.11	&	99.50	&	99.53	&	\textbf{100}	&	99.61	&	99.91	&	99.80	& 	96.91	&	99.71	&	96.05 \\
6	&	95.10	&	95.10	&	\textbf{99.30}	&	95.80	&	97.89	&	97.90	&	96.47	& 	94.12	&	95.08	&	96.15 \\
7	&	75.00	&	\textbf{92.35}	&	72.85	&	80.88	&	91.25	&	84.70	&	84.58	& 	79.95	&	89.90	&	80.25 \\
8	&	40.17	&	53.41	&	43.02	&	44.54	&	75.92	&	46.59	&	73.09	& 	68.19	&	70.52	&	\textbf{77.78} \\
9	&	74.88	&	79.51	&	71.01	&	84.32	&	83.20	&	78.41	&	78.14	& 	76.39	&	54.89	&	\textbf{84.66} \\
10	&	51.64	&	53.57	&	47.20	&	49.61	&	48.80	&	53.15	&	58.01	& 	48.10	&	49.89	&	\textbf{64.63} \\
11	&	78.37	&	66.22	&	76.66	&	83.59	&	69.21	&	78.91	&	73.35	& 	74.64	&	77.36	&	\textbf{88.61} \\
12	&	68.40	&	84.44	&	68.78	&	75.12	&	85.55	&	89.24	&	86.74	&	85.49	&	60.46	&	\textbf{98.57} \\
13	&	69.47	&	68.41	&	43.16	&	60.70	&	90.85	&	79.65	&	\textbf{91.16}	& 	88.58	&	81.71	&	83.09 \\
14	&	\textbf{100}	&	\textbf{100}	&	88.66	&	\textbf{100}	&	90.66	&	\textbf{100}	&	\textbf{100}	&	99.19	&	95.14	&	99.72 \\
15	&	98.10	&	99.58	&	\textbf{99.79}	&	99.48	&	76.83	&	98.59	&	77.73	& 	96.40	&	65.35	&	81.90 \\
\hline
OA	&	76.88	&	79.58	&	77.12	&	79.98	&	82.23	&	81.38	&	83.42	& 	82.34	&	78.16	&	\textbf{84.56} \\
AA	&	80.29	&	80.78	&	78.86	&	82.21	&	84.08	&	84.44	&	84.17	&	84.43	&	79.98	&	\textbf{86.18} \\
Kappa	&	0.7513	&	0.7815	&	0.7528	&	0.7824	&	0.8013	&	0.7991	&	0.8114	&	0.8052	&	0.7616	&	\textbf{0.8328} \\

\hline
\hline
\end{tabular}
\label{Houston_classification_results}
\end{table*}

\subsection{Compared Methods}

\begin{figure*}[tp]
\centering
\subfigure[SVM]{\includegraphics[width=29mm]{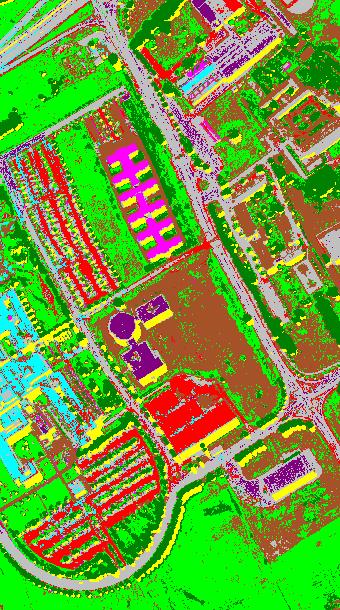}}
\hspace{+1pt}
\subfigure[EMP]{\includegraphics[width=29mm]{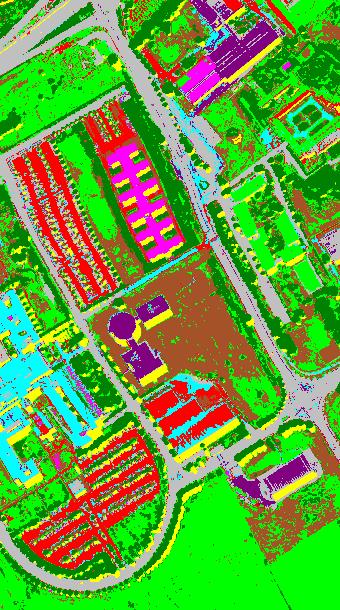}}
\hspace{+1pt}
\subfigure[JSR]{\includegraphics[width=29mm]{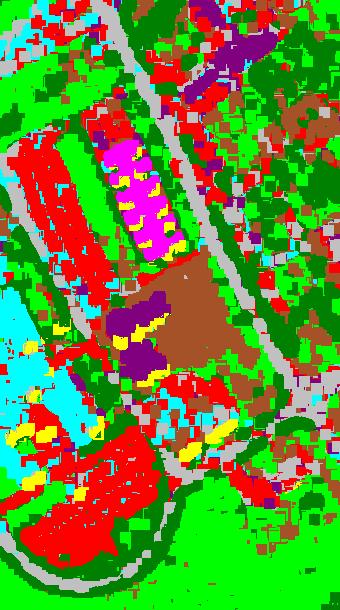}}
\hspace{+1pt}
\subfigure[EPF]{\includegraphics[width=29mm]{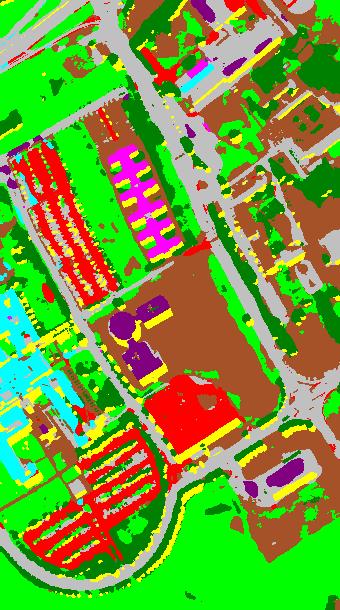}}
\hspace{+1pt}
\subfigure[3D-CNN]{\includegraphics[width=28mm]{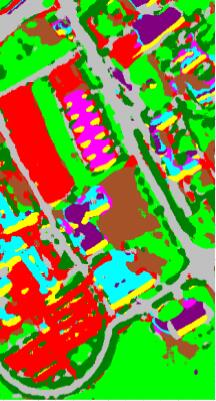}}
\hspace{+1pt}
\subfigure[CNN-PPF]{\includegraphics[width=29mm]{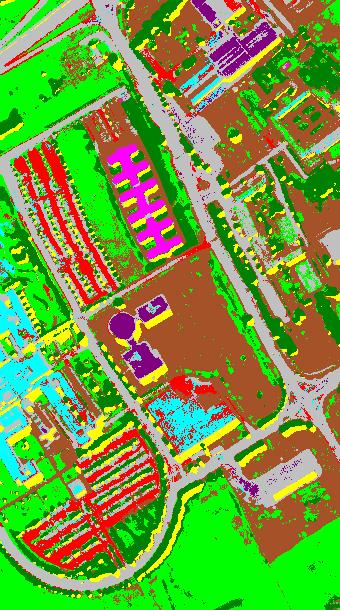}}
\hspace{+1pt}
\subfigure[Gabor-CNN]{\includegraphics[width=28mm]{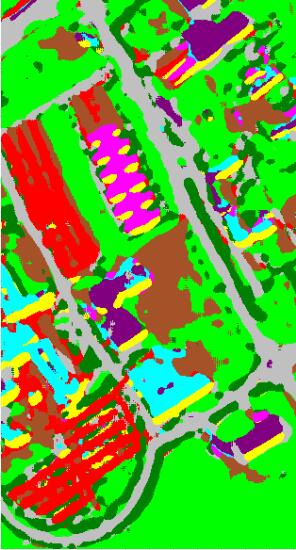}}
\hspace{+1pt}
\subfigure[S-CNN]{\includegraphics[width=29mm]{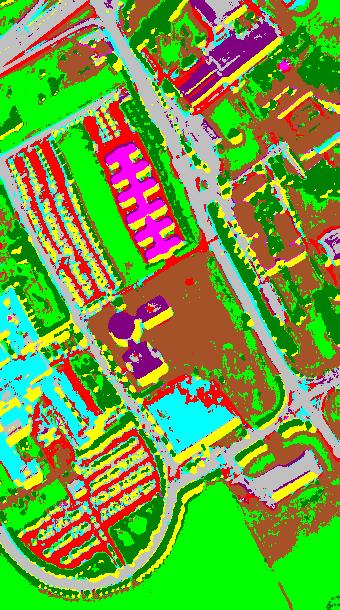}}
\hspace{+1pt}
\subfigure[3D-GAN]{\includegraphics[width=28mm]{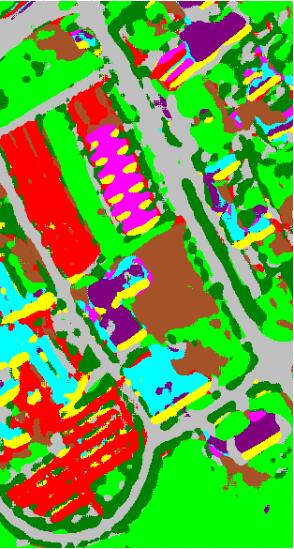}}
\hspace{+1pt}
\subfigure[DFFN]{\includegraphics[width=29mm]{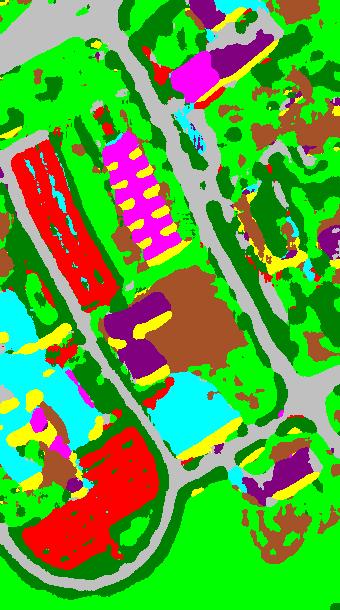}}

\caption{Classification maps for the University of Pavia data set obtained by (a) SVM \cite{SVM}, (b) EMP \cite{EMP1}, (c) JSR \cite{JSR}, (d) EPF \cite{EPF}, (e) 3D-CNN \cite{Chen-CNN}, (f) CNN-PPF \cite{Li-CNN-DPPF}, (g) Gabor-CNN \cite{Chen-Gabor}, (h) S-CNN \cite{Liu-simase-CNN}, (i) 3D-GAN \cite{Chen-GAN-HSI}, and (j) DFFN \cite{DFFN}.}
\label{paviau_classification_maps}
\end{figure*}

\begin{figure*}[tp]
\centering
\subfigure[SVM]{\includegraphics[width=28mm]{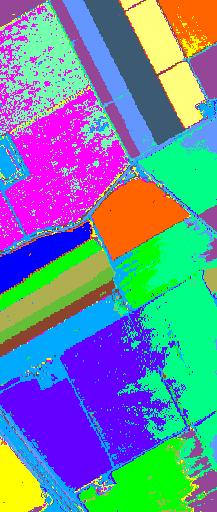}}
\hspace{+2pt}
\subfigure[EMP]{\includegraphics[width=28mm]{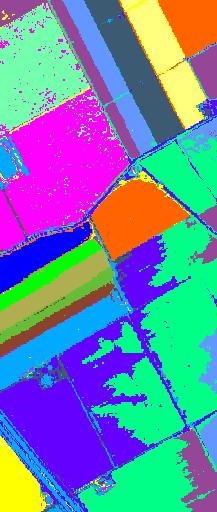}}
\hspace{+2pt}
\subfigure[JSR]{\includegraphics[width=28mm]{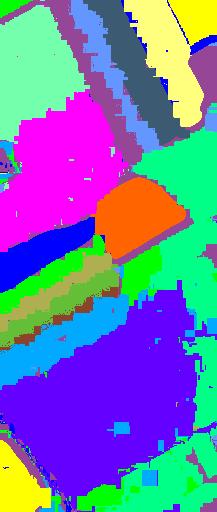}}
\hspace{+2pt}
\subfigure[EPF]{\includegraphics[width=28mm]{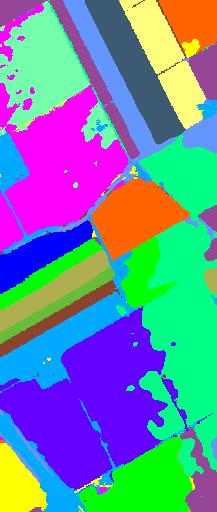}}
\hspace{+2pt}
\subfigure[3D-CNN]{\includegraphics[width=28mm]{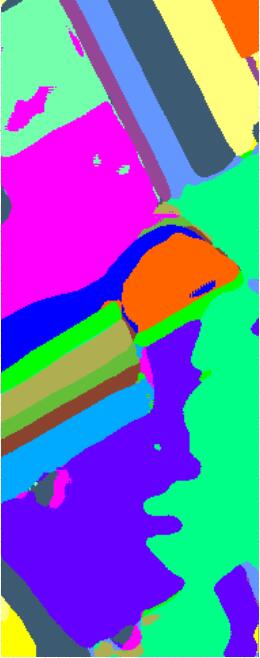}}
\hspace{+2pt}
\subfigure[CNN-PPF]{\includegraphics[width=28mm]{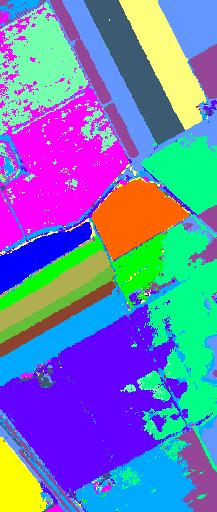}}
\hspace{+2pt}
\subfigure[Gabor-CNN]{\includegraphics[width=28mm]{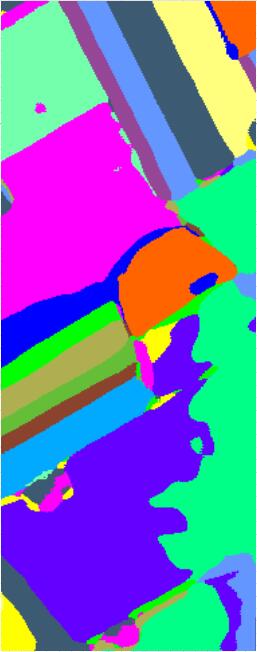}}
\hspace{+2pt}
\subfigure[S-CNN]{\includegraphics[width=28mm]{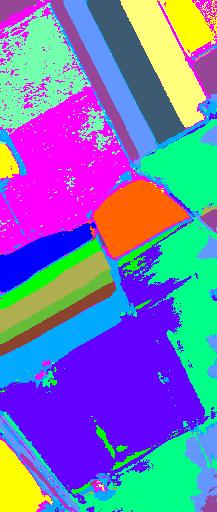}}
\hspace{+2pt}
\subfigure[3D-GAN]{\includegraphics[width=28mm]{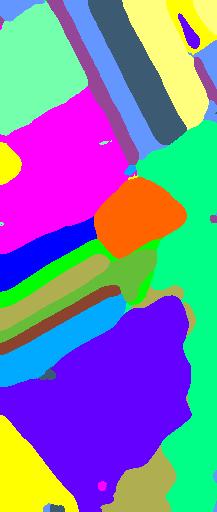}}
\hspace{+2pt}
\subfigure[DFFN]{\includegraphics[width=28mm]{./DFFN_salinas_full.jpg}}

\caption{Classification maps for the Salinas data set obtained by (a) SVM \cite{SVM}, (b) EMP \cite{EMP1}, (c) JSR \cite{JSR}, (d) EPF \cite{EPF}, (e) 3D-CNN \cite{Chen-CNN}, (f) CNN-PPF \cite{Li-CNN-DPPF}, (g) Gabor-CNN \cite{Chen-Gabor}, (h) S-CNN \cite{Liu-simase-CNN}, (i) 3D-GAN \cite{Chen-GAN-HSI}, and (j) DFFN \cite{DFFN}.}
\label{salinas_classification_maps}
\end{figure*}

In this review, we have investigated several recent state-of-the-art deep learning-based approaches, including 3D-CNN \cite{Chen-CNN}, Gabor-CNN \cite{Chen-Gabor}, CNN with pixel-pair features (CNN-PPF) \cite{Li-CNN-DPPF}, siamese CNN (S-CNN) \cite{Liu-simase-CNN}, 3D-GAN \cite{Chen-GAN-HSI}, and the deep feature fusion network (DFFN) \cite{DFFN}, for HSI classification. Specifically, the 3D-CNN exploits 3-D convolutional filters to directly extract spectral-spatial features from the original hyperspectral cube. However, the network architecture adopted in 3D-CNN is relatively simple and the correlation between different layers is neglected. The DFFN adopts the DRN \cite{DRN}, which can be considered a more powerful network, to extract more discriminative features. In addition, the features from different-level layers are further fused to explore the correlation between layers. The main drawback of DFFN is that the optimal feature fusion mechanism depends on a hand-crafted setting with abundant experiments. In the Gabor-CNN, the Gabor filtering is first utilized as a preprocessing technique to extract spatial features of HSIs. Then, the filtered features are fed into a simple CNN-based classifier. Instead of considering a pixel-wised semantic information, CNN-PPF and S-CNN focus on exploring the correlation between samples. In more detail, CNN-PPF extracts the pixel-pair features via CNN. However, the convolutional operation is mainly conducted in the spectral domain, and thus, the spatial information is not considered for CNN-PPF. By contrast, S-CNN adopts a two-branch CNN to simultaneously extract spectral-spatial features of HSIs. But, in such a method, the computational cost may be huge due to the high dimension vector in the Euclidean space. In addition, 3D-GAN utilizes adversarial training to improve the generalization capability of the discriminative CNN, which is very useful when the number of training samples is limited.  \par
Apart from the above deep learning-based methods, some classical classification techniques, including the SVM \cite{SVM}, EMP \cite{EMP1}, joint spare representation (JSR) \cite{JSR}, and edge-preserving filtering (EPF) \cite{EPF} are also considered for comparison. As the spectral feature-based method, the SVM was implemented in the library for support vector machines (LIBSVM) library \cite{LIBSVM}, where the Gaussian kernel with fivefold cross validation was adopted for this classifier. The other three methods are all spectral-spatial feature-based methods. Specifically, for the EMP, the morphological profiles were constructed with the first three principal components. Furthermore, a circular structural element, a step size increment of two, and four openings and closings were performed for each principal component. For the JSR, the spatial information within a fixed-size local region was utilized by the joint sparse regularization. The EPF method was implemented by using the code which is available on Dr. Xudong Kang's homepage\footnote{http://xudongkang.weebly.com}.  \par
Among these investigated methods, the SVM and CNN-PPF have only utilized spectral features during the classification. In contrast, the rest of methods, including the EMP, JSR, EPF, 3D-CNN, Gabor-CNN, S-CNN, 3D-GAN, and DFFN, are belong to classification methods based on spectral-spatial features.

\begin{table*}
\centering
\footnotesize
  \caption{The Classification Accuracies (in Percentages) Obtained by SVM \cite{SVM}, EMP \cite{EMP1}, EPF \cite{EPF},  JSR \cite{JSR}, 3D-CNN \cite{Chen-CNN}, Gabor-CNN \cite{Chen-Gabor}, CNN-PPF \cite{Li-CNN-DPPF}, S-CNN \cite{Liu-simase-CNN}, 3D-GAN \cite{Chen-GAN-HSI}, and DFFN \cite{DFFN} on the University of Pavia Image. The Best Accuracies Are Marked in Bold.}
\begin{tabular}{ccccccccccc}
\hline
\hline

classes	&	SVM	&	EMP	&	JSR	&	EPF	&	3D-CNN	&	CNN-PPF	&	Gabor-CNN	&	S-CNN	&	3D-GAN	&	DFFN \\
\hline
1	&	85.49	&	98.70	&	83.80	&	96.08	&	99.03	&	97.23	&	\textbf{99.53}	&	95.47	&	99.18	&	\textbf{99.53} \\
2	&	92.12	&	\textbf{98.92}	&	96.31	&	98.24	&	98.11	&	95.27	&	98.21	&	98.71	&	98.86	&	97.71 \\
3	&	85.77	&	94.75	&	98.63	&	95.25	&	88.56	&	95.13	&	89.74	&	97.32	&	94.94	&	\textbf{99.89} \\
4	&	96.41	&	96.66	&	93.89	&	\textbf{98.27}	&	83.51	&	96.89	&	93.02	&	97.72	&	90.15	&	97.88 \\
5	&	98.60	&	98.87	&	98.38	&	\textbf{100}	&	99.49	&	99.99	&	99.42	&	\textbf{100}	&	99.49	&	99.48 \\
6	&	92.52	&	84.52	&	99.54	&	98.16	&	95.33	&	98.55	&	98.77	&	97.67	&	98.56	&	\textbf{99.69} \\
7	&	93.79	&	87.97	&	98.50	&	98.79	&	96.31	&	96.56	&	98.82	&	98.36	&	92.74	&	\textbf{100} \\
8	&	86.56	&	\textbf{98.72}	&	96.51	&	96.81	&	97.58	&	94.43	&	94.12	&	95.65	&	97.18	&	98.59 \\
9	&	97.97	&	99.87	&	75.88	&	97.91	&	96.25	&	99.39	&	97.91	&	\textbf{100}	&	98.51	&	99.61 \\
\hline
OA	&	90.78	&	96.16	&	94.43	&	97.58	&	96.37	&	97.63	&	97.33	&	97.93	&	97.81	&	\textbf{98.57} \\
AA	&	92.14	&	95.08	&	92.57	&	97.72	&	94.82	&	97.04	&	96.62	&	97.88	&	96.65	&	\textbf{99.16} \\
Kappa	&	0.8813	&	0.9504	&	0.9349	&	0.9688	&	0.9502	&	0.969	&	0.9662	&	0.9743	&	0.9697	&	\textbf{0.9808} \\

\hline
\hline
\end{tabular}
\label{paviau_classification_results}
\end{table*}

\subsection{Classification Results}

\begin{table*}
\centering
\footnotesize
  \caption{The Classification Accuracies (in Percentages) Obtained by SVM \cite{SVM}, EMP \cite{EMP1}, EPF \cite{EPF},  JSR \cite{JSR}, 3D-CNN \cite{Chen-CNN}, Gabor-CNN \cite{Chen-Gabor}, CNN-PPF \cite{Li-CNN-DPPF}, S-CNN \cite{Liu-simase-CNN}, 3D-GAN \cite{Chen-GAN-HSI}, and DFFN \cite{DFFN} on the Salinas Image. The Best Accuracies Are Marked in Bold.}
\begin{tabular}{ccccccccccc}
\hline
\hline

classes	&	SVM	&	EMP	&	JSR	&	EPF	&	3D-CNN	&	CNN-PPF	&	Gabor-CNN	&	S-CNN	&	3D-GAN	&	DFFN \\
\hline
1	&	99.61	&	98.77	&	\textbf{100}	&	\textbf{100}	&	99.94	&	99.84	&	\textbf{100}	&	\textbf{100}	&	75.35	&	99.99 \\
2	&	99.69	&	98.27	&	99.83	&	98.98	&	85.45	&	99.77	&	88.06	&	97.75	&	98.49	&	\textbf{99.94} \\
3	&	99.56	&	99.74	&	99.92	&	99.99	&	\textbf{100}	&	98.11	&	99.25	&	98.88	&	\textbf{100}	&	\textbf{100} \\
4	&	99.41	&	98.98	&	94.77	&	\textbf{100}	&	99.77	&	99.57	&	\textbf{100}	&	\textbf{100}	&	99.76	&	\textbf{100} \\
5	&	98.72	&	97.33	&	83.31	&	99.49	&	99.96	&	98.54	&	98.88	& 	99.93	&	\textbf{100}	&	99.11 \\
6	&	99.77	&	99.46	&	92.34	&	\textbf{100}	&	\textbf{100}	&	99.92	&	\textbf{100}	&	89.48	&	99.97	&	99.95 \\
7	&	99.52	&	99.11	&	95.92	&	99.93	&	99.60	&	\textbf{99.96}	&	98.94	&	99.30	&	99.45	&	99.43 \\
8	&	76.74	&	93.79	&	96.97	&	92.55	&	99.31	&	89.11	&	99.48	&	98.69	&	98.30	&	\textbf{99.56} \\
9	&	99.37	&	99.73	&	99.99	&	99.99	&	99.97	&	99.69	&	97.27	&	99.34	&	99.95	&	\textbf{100} \\
10	&	95.13	&	98.38	&	92.42	&	99.69	&	99.41	&	97.78	&	99.21	&	\textbf{100}	&	99.79	&	99.79 \\
11	&	99.32	&	98.69	&	93.86	&	\textbf{100}	&	\textbf{100}	&	99.33	&	\textbf{100}	&	99.93	&	\textbf{100}	&	99.48 \\
12	&	99.70	&	96.52	&	83.94	&	99,53	&	\textbf{100}	&	\textbf{100}	&	\textbf{100}	&	99.89	&	\textbf{100}	&	99.84 \\
13	&	99.15	&	99.37	&	89.25	&	99.99	&	\textbf{100}	&	99.67	&	\textbf{100}	&	88.62	&	\textbf{100}	&	99.96 \\
14	&	98.38	&	98.77	&	70.06	&	99.94	&	\textbf{100}	&	98.75	&	99.79	&	88.72	&	\textbf{100}	&	99.95 \\
15	&	75.56	&	92.75	&	96.74	&	92.71	&	90.36	&	89.99	&	94.31	&	90.62	&	\textbf{99.52}	&	99.45 \\
16	&	99.23	&	97.68	&	99.81	&	\textbf{99.96}	&	85.93	&	99.07	&	93.38	&	\textbf{99.96}	&	90.25	&	\textbf{99.96} \\
\hline
OA	&	90.92	&	96.89	&	95.33	&	97.32	&	97.28	&	94.87	&	97.63	& 	97.62	&	98.22	&	\textbf{99.71} \\
AA	&	96.18	&	97.95	&	94.78	&	98.88	&	97.48	&	98.07	&	98.04	&	96.94	&	97.75	&	\textbf{99.78} \\
Kappa	&	0.8985	&	0.9707	&	0.9307	&	0.9700	&	0.9695	&	0.9404	&	0.9734	&	0.9510	&	0.9793	&	\textbf{0.9967} \\

\hline
\hline
\end{tabular}
\label{salinas_classification_results}
\end{table*}

The first experiment was performed on the Houston data set. In this experiment, the training samples were given according to the 2013 GRSS data fusion contest. The amount of training and test samples per class is shown in Table I. Fig. \ref{Houston_classification_maps} shows the classification maps obtained by different methods. From this figure, we can see that the classification maps obtained by the SVM and JSR methods are not very satisfactory since some noisy estimations are still visible. By contrast, other methods perform much better in removing ``noisy pixels" and deliver a smoother appearance in their classification results. By comparing two filtering-based methods, i.e., EPF and Gabor-CNN, we can see that the classification map of EPF seems to be over-smoothing, but the Gabor-CNN preserves more details in edges. Apart from visual comparison, Table \ref{Houston_classification_results} gives quantitative results of various methods on the image, where three metrics, i.e., overall accuracy (OA), average accuracy (AA), and Kappa coefficient, are adopted to evaluate the classification performance. All the classification accuracy values reported in Table \ref{Houston_classification_results} are the average results over ten experiments. As can be seen, except for 3D-GAN, the deep learning-based methods obtain satisfactory classification accuracies (e.g., all OAs are above 80\%), which is overall higher than traditional methods, including the SVM, EMP, JSR and EPF. Among these deep learning-based methods, CNN-PPF which only considers spectral information of HSIs obtains the inferior classification results compared with other spectral-spatial features methods, e.g., 3D-CNN, Gabor-CNN, S-CNN, and DFFN. Moreover, the DFFN which combines the residual learning with feature fusion in the a deep CNN framework delivers the best classification accuracies in terms of OA, AA, and Kappa.   \par

\begin{figure*}
\begin{center}
\subfigure[]{\includegraphics[width=170mm]{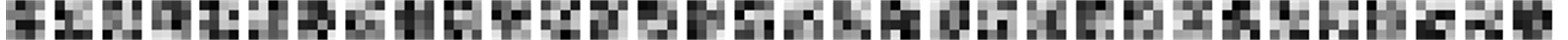}}
\subfigure[]{\includegraphics[width=170mm]{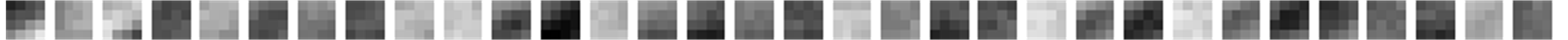}}
\end{center}
\caption{Weights of the first convolutional layer. The size of each convolutional kernel is $4\times4$. Each kernel of 32 kernels in the first convolutional layer is shown in the forms of a tiny image. (a) Randomly initialized weights of the first convolution layer on the Salinas data set. (b) Learned weights of the first convolution layer on the Salinas data set. }
\label{weights}
\end{figure*}

\begin{figure*}
\begin{center}
\includegraphics[width=160mm]{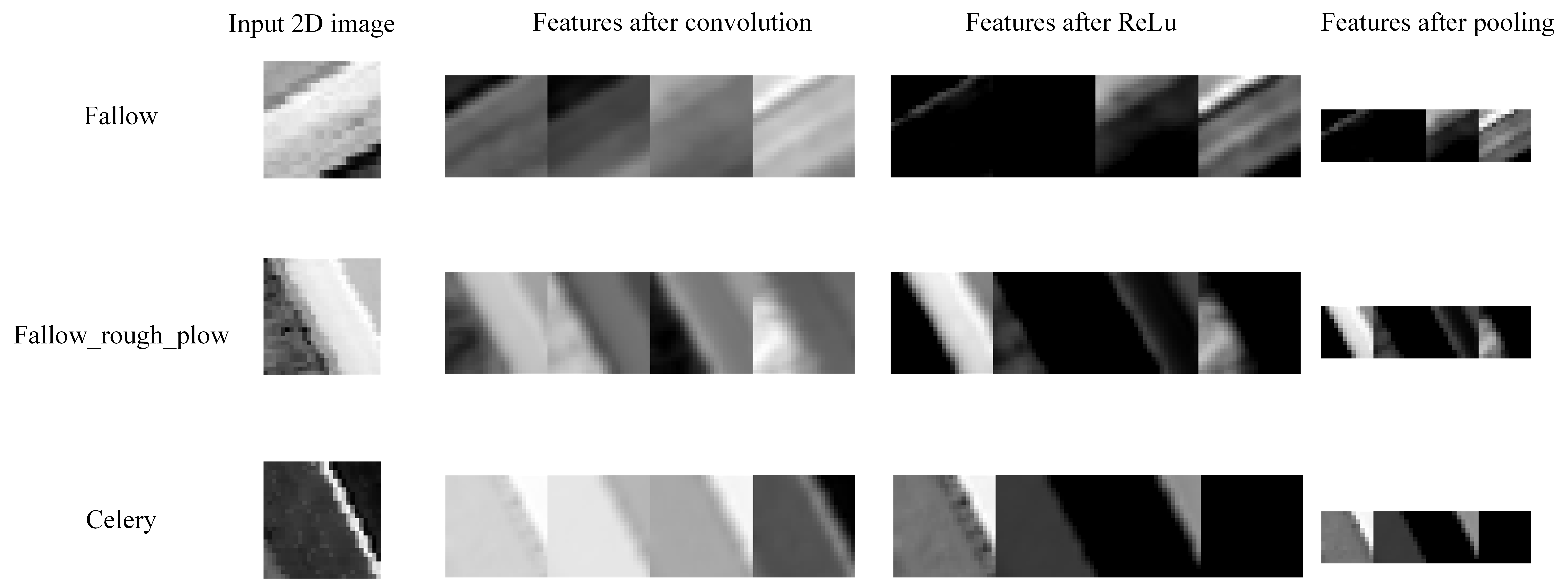}
\end{center}
\caption{Extracted features of the Salinas data set. Each row of the images represents one class. There are four columns in the figure. The first column represents the input images. The second column shows the four feature maps after the first convolution. The third column shows the four feature maps after the first ReLU operation. The last column shows the four features of the first pooling operation. }
\label{features}
\end{figure*}

The second and third experiments were conducted on the University of Pavia and Salinas images. As mentioned above, 200 samples per class were randomly selected as the training samples and the rest of samples as the test samples (see Tables II and III). The selected training samples account for about 4\% and 6\% of the whole labeled reference data for the University of Pavia and Salinas images, respectively, which provide a challenging test set. Figs. \ref{paviau_classification_maps} and \ref{salinas_classification_maps} show the classification maps obtained by different methods. Tables \ref{paviau_classification_results} and \ref{salinas_classification_results} present the quantitative results (averaged over ten experiments) of different methods. \par

From the above experimental results, the deep learning-based methods show great advantages over other traditional methods in terms of visual classification maps and quantitative results. For instance, considering that in comparison of several SVM-based classifiers, including SVM, EMP, EPF, and S-CNN, S-CNN performs the best on three hyperspectral data sets, which can be used to verify the effectiveness of deep features compared with hand-crafted features. In addition, for two filter-based methods, i.e., EPF and Gobar-CNN, the OA of Gabor-CNN is about 3.5\% higher than that of EPF on the Houston image, which demonstrates that combining the filter technique with deep leaning can give good classification results. Although CNN-PPF only utilizes spectral information, it still delivers a better performance than what was obtained by the other four traditional-based methods on the Houston and University of Pavia images.

\subsection{Deep Feature Visualization}

In general, deep learning can be regarded as a black box in most applications, where the inside information of network is often unclear. Actually, exploring the inside features is very useful for analyzing the network performance and further designing the deep architecture. In this section, we used Salinas data set as an example to visualize the deep features. The weights of different convolutional kernels in the first convolutional layer are shown in Fig. \ref{weights}. The Fig. \ref{weights} (a) is random initial weights, and Fig. \ref{weights} (b) demonstrates the learned weights. From this figure, we can see that the distribution of the weights for each filter become more regular and present evident textural features after training. For example, the intensities of the learned weights in the first kernel are low on the left side and high on the right side, \emph{i.e.}, the weights reveal a trend of gradual decrease, which is different from the structure of randomly initialized weights significantly.  \par
\begin{figure}
\begin{center}
\includegraphics[width=80mm]{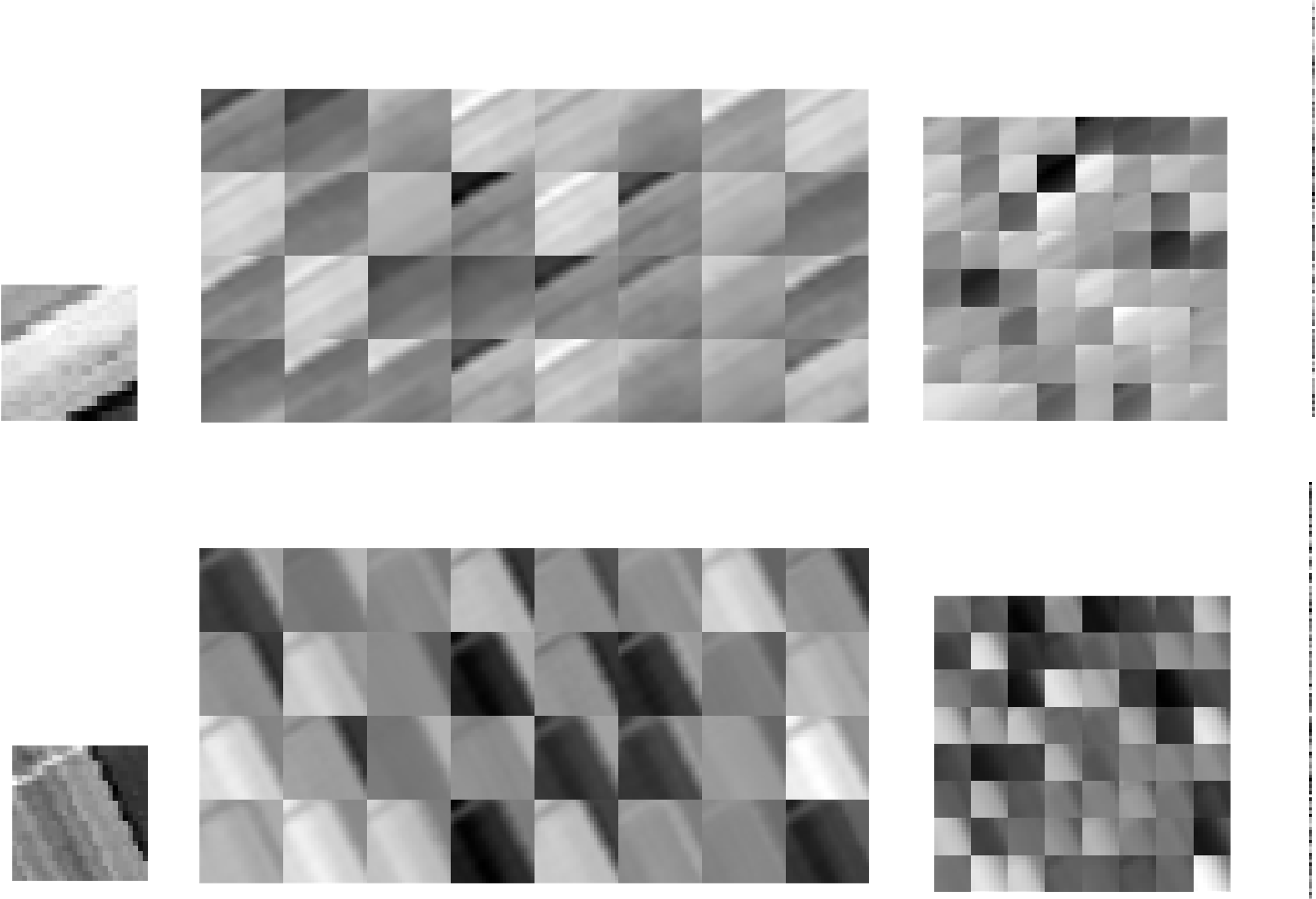}
\end{center}
\caption{ Extracted features after three convolutional layers on Fallow and Fallow\_rough\_plow samples. The first column represents the $27\times27$ input images; the second column shows 32 feature maps after the first convolution layer, for which the size of each feature map is $24\times24$; the third column shows 64 feature maps after the second convolution layer, for which the size of each feature map is $8\times8$; and the number of feature maps after the third convolution layer shown in the last column is 128, and the size of each feature map is $1\times1$. }
\label{features_classes}
\end{figure}

In addition, Fig. \ref{features} shows the features after the convolutional layer, ReLU layer, and pooling layer. Fig. \ref{features_classes} shows the extracted features after three convolutional layers of Fallow and Fallow\_roughg\_plow samples. From Figs. \ref{features} and \ref{features_classes}, one can see that the earlier convolution layers can extract some simple features like texture and edge information, and those lower level features can be composed to become more complicated high-level features though the deeper convolution layers. The learning process is totally automatic, which makes CNNs more suitable for coping with varieties of situations. \par

\subsection{Effectiveness Analysis of Strategies for Limited Samples}

In this section, we will validate the effectiveness of some strategies proposed in Section IV for the problem of limited training samples. Specifically, we build a simple CNN and adopt three strategies, i.e., data augmentation, transfer learning and residual learning, to improve the classification accuracy. These methods, i.e., original CNN, CNN with data augmentation, CNN with transfer learning, and CNN with residual learning, are denoted as CNN-Original, CNN-DA, CNN-TL, and CNN-RL, respectively. The following experiments were conducted in the Salinas data set, where the training set consists of 5, 10, 15, 20, 25, and 30 labeled samples per class, respectively, and the remaining  available sample were regarded as test data sets. All the used CNNs have the same architectures which consist of seven convolutional layers followed by the batch normalization operation, one global pooling layer, and two fully connected layers. The detailed structural information is presented in Table \ref{CNN-struture}. For the CNN-DA, new samples were generated according to the equation (14). For the CNN-TL, the CNN was pretrained on the Indian Pines data set because the Indian Pines and Salinas images were collected by the same sensor, i.e., the Airborne Visible/Infrared Imaging Spectrometer sensor. For CNN-RL, the residual learning technique was incorporated into the CNN to optimize the network. \par

\begin{figure}
\begin{center}
\includegraphics[width=68mm]{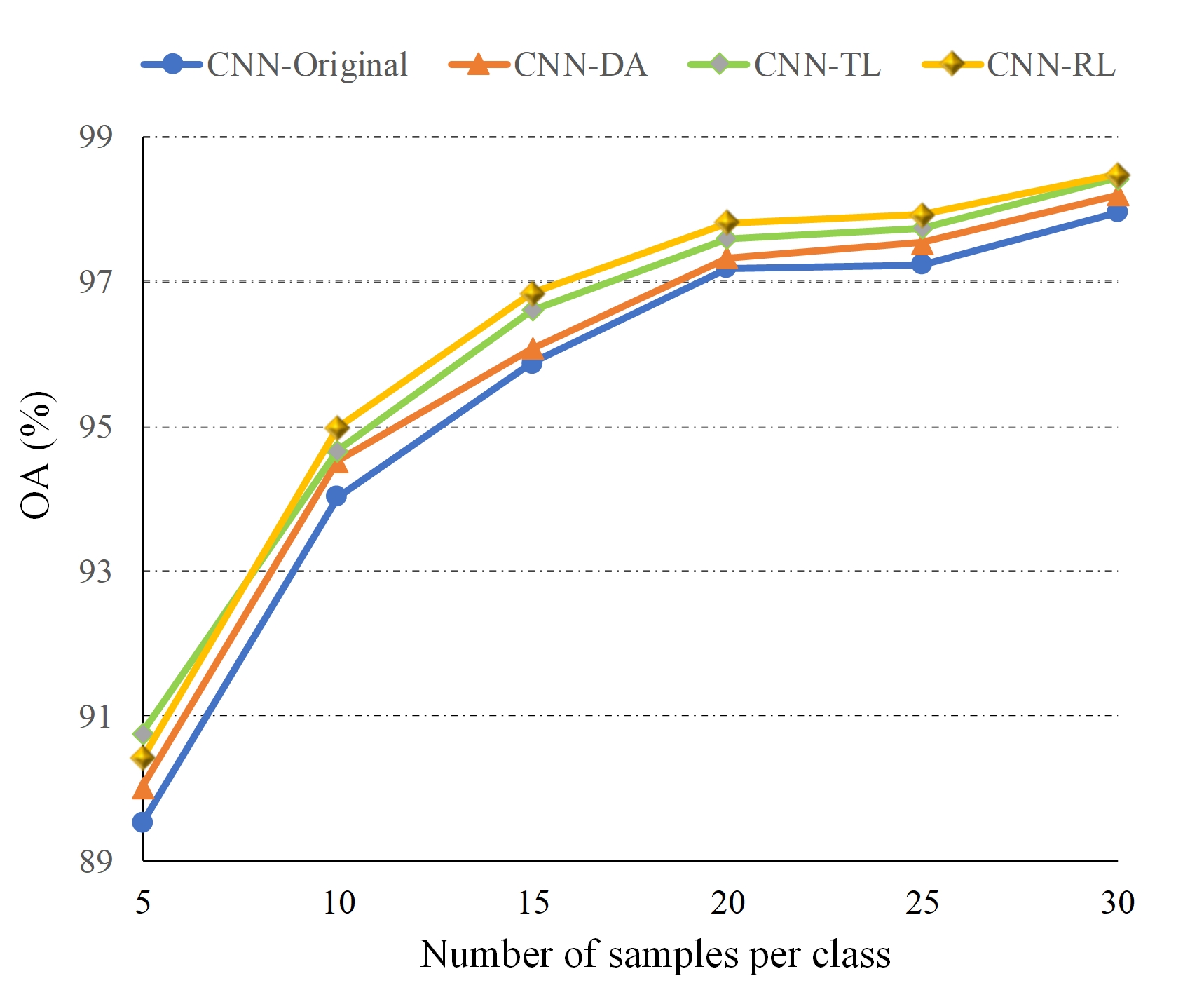}
\end{center}
\caption{ The OA values obtained by different methods versus number of samples. }
\label{limited samples}
\end{figure}

Fig. \ref{limited samples} shows the classification accuracies obtained by different methods in terms of OA values. From this figure, one can see that OA values of CNN-DA, CNN-TL, and CNN-RL are higher than OA values of CNN-Original for all training situations, which demonstrates that these strategies, indeed, improve the network performance to some extent when few training samples are available. The improvement almost approaches 1\% by comparing CNN-TL with CNN-Original when the training set consists of 5 samples per class. In addition, CNN-RL performs better than other methods for most situations, which demonstrates that the residual learning is a very useful network optimization method for HSI classification.  \par

\begin{table}
\centering
\footnotesize
  \caption{The Configuration of CNN Model }
  \label{CNN-struture}
\begin{tabular}{cccc}
\hline
\hline
Layer No.   & Type          & Size  & Feature maps  \\
\hline
1-3          & Convolution       & $3\times{3}$                & 16  \\
4-5          & Convolution       & $3\times{3}$                & 32  \\
6-7          & Convolution       & $3\times{3}$                & 64  \\
8            & Pooling       & $3\times{3}$                & 64  \\
9        & Fully connected            & -              & 200  \\
10       & Fully connected        & -      & 16 \\
\hline
\hline
\end{tabular}
\end{table}

\section{Conclusion}
Recently, deep learning-based HSI classification has drawn a significant attention in the remote sensing field and obtained good performance. In contrast to traditional hand-crafted feature-based classification methods, deep learning can automatically learn complex features of HSIs with a large number of hierarchical layers. In this literature survey, we briefly introduced several deep models that are often used to classify HSIs, including SAE, DBN, CNN, RNN, and GAN. Then, we focused on deep learning-based classification methodologies for HSIs and provided a general and comprehensive overview on the existing methods in a unified framework. Specifically, these deep networks used in the HSI classification were divided into spectral-feature networks, spatial-feature networks, and spectral-spatial-feature networks, where each category extracts the corresponding feature. Through this framework, we can easily see that deep networks make full use of different feature types for classification. We have also compared and analyzed the performances of various HSI classification methods, including four traditional machine learning-based methods and six deep learning-based methods. The classification accuracies obtained by different methods demonstrate that deep learning-based methods overall outperform the non-deep-learning-based methods and the DFFN which combines the residual learning and feature fusion achieves best classification performance. Furthermore, deep features and network weights were visualized, which is useful for analyzing the network performance and further designing the deep architecture. In addition, considering the fact that available training samples in remote sensing are usually very limited and training deep network requires a large number of samples, we also included some strategies to improve classification performance. We have also conducted experiments to verify and compared the effectiveness of these strategies. The final results show that the residual learning obtains the highest improvement among all approaches. This experimental result may provide some guidelines for the future study on this topic.

\ifCLASSOPTIONcaptionsoff
  \newpage
\fi




%





\bibliographystyle{IEEEbib}
\bibliography{refs}

\end{document}